\documentclass[11pt]{article}

\usepackage{cite}
\usepackage{amsfonts}
\usepackage{amsmath}
\usepackage{wrapfig}
\usepackage{latexsym}
\usepackage{fancybox}
\usepackage{bm}
\usepackage{accents}
\usepackage{float}
\usepackage{mathtools}

\makeatletter
  \def\widebar{\accentset{{\cc@style\underline{\mskip10mu}}}}
  \def\Widebar{\accentset{{\cc@style\underline{\mskip8mu}}}}
  \makeatother

\numberwithin{equation}{section}
\allowdisplaybreaks

\def\spa#1{\phantom{\fbox{\rule[-#1cm]{0cm}{0cm}}}}

\def\[#1\]{\begin{align}#1\end{align}}
\newcommand{\vev}[1]{\left\langle #1 \right\rangle}
\newcommand{\nn}{\nonumber}
\newcommand{\wt}[1]{\widetilde{#1}}
\newcommand{\wh}[1]{\widehat{#1}}

\newcommand{\bn}{\bm\nabla}
\newcommand{\thbar}{\tilde{\hbar}}
\newcommand{\param}{\frac{\omega}{mc_s^2}}
\newcommand{\ve}{v_{\text{eff}}}
\newcommand{\rhoin}{\rho_{\text{in}}}
\newcommand{\vac}{|\psi_0\rangle_{\text{in}}}
\newcommand{\tvac}{{}_{\text{in}}\langle\psi_0|}
\newcommand{\tr}{\text{tr}}
\newcommand{\bra}[1]{\left\langle#1\right|}
\newcommand{\ket}[1]{\left|#1\right\rangle}
\newcommand{\C}[2]{{}_{#1}C_{#2}}

\begin{document}

\hfuzz=100pt
\title{{\Large \bf{
Decrease of the entanglement entropy of the Hawking radiation 
induced by backreaction in the Bose-Einstein condensate
}}}
\author{Tsunehide Kuroki$^{a}$\footnote{kuroki@toyota-ti.ac.jp}  
  \spa{0.5} \\
\\
$^a${\small{\it Theoretical Physics Laboratory, Toyota Technological Institute}}
\\ {\small{\it 2-12-1 Hisakata, Tempaku-ku, Nagoya 468-8511, Japan}} 
\spa{0.3} 
}
\date{}

\maketitle
\centerline{}

\begin{abstract} 
We analytically study the effect of backreaction from analog Hawking radiation on its entanglement entropy in the Bose-Einstein condensate (BEC). The backreaction is expected to play an essential role in the decrease of the entanglement entropy and in realizing the Page curve. Since the BEC theory has microscopic Hamiltonian and thus exhibits unitarity, it is desirable to reproduce the Page curve explicitly by using the Hamiltonian. As the first step towards this goal, we concretely study the BEC with a step-like configuration that has been extensively studied in the literature. By using the microscopic theory, we derive an explicit form of backreaction from analog Hawking radiation. Combining it with the known results of the Bogoliubov coefficients, we analytically compute the entanglement entropy of the Hawking radiation, and show that  the backreaction reduces the growth of the entanglement entropy as expected for sufficiently low energy modes over a wide range of the parameter characterizing the step-like configuration. 
\end{abstract}

\renewcommand{\thefootnote}{\arabic{footnote}}
\setcounter{footnote}{0}

\newpage

\section{Introduction}
\label{sec:introduction}

One of the central challenges in theoretical physics is the resolution of information loss problem 
\cite{Hawking:1976ra}. Attempts to resolve it will provide insights into fundamental degrees of freedom in quantum gravity. We expect that Hawking radiation plays a role analogous to that of blackbody radiation in the old quantum theory. So far, several ideas for resolving the problem have been proposed, such as the island formula \cite{Almheiri:2020cfm} and the non-separability of the Hilbert space between the inside and outside of a black hole (BH) \cite{Geng:2026asi}.\footnote{Recent studies suggest that these perspectives are not in tension, but can be understood in a unified framework, see e.g. \cite{Geng:2023zhq,Geng:2025rov}. We thank Hao Geng for bringing this to our attention.} We have not, however, reached a unified understanding yet, partly for the lack of microscopic formulation of quantum gravity. 

In view of this situation, it would be invaluable to investigate 
a similar physical system with microscopic description. This is where the analog gravity approach based on the Bose-Einstein condensate (BEC) comes into play.  It is possible there to construct a configuration which  
mimics a BH and is indeed known to exhibit the analog of the Hawking radiation as well \cite{Unruh:1980cg}. We can arrange it in such a way that the condensate has one-dimensional flow 
and its velocity exceeds sound velocity only in restricted region, for example, $x<0$\cite{Macher:2009tw,Recati:2009ya}. This configuration realizes the analog BH from which the sound cannot escape. In this case we have a definite Hamiltonian and manifest unitarity. Thus, if the initial state is pure, the entanglement entropy of the analog Hawking radiation should follow the so-called Page curve \cite{Page:1993df}, and it should be possible in principle to analyze how it is realized by using the microscopic Hamiltonian. We note that in the case of an analog BH, the Bekenstein-Hawking entropy is not available. The resolution of the information loss problem boils down to identifying where the original argument by Hawking \cite{Hawking:1976ra} undergoes significant changes. In the BEC case, it is probable that the backreaction of the Hawking radiation which is neglected in the Hawking's original argument plays a crucial role. Although it is not clear that this is also the case with the gravity, it is intriguing and invaluable in itself to take account of it and see what happens to the entanglement entropy. 

In order to reproduce the Page curve, our expectation is that the entanglement entropy would decrease due to backreaction. The aim of this paper is to check that it indeed reduces the growth of the entanglement entropy in a concrete setup.\footnote{Backreaction in analog gravity itself has been discussed in various contexts, see e.g. \cite{Schutzhold:2005ex,Fischer:2005iy,Liberati:2019fse,Patrick:2019kis,Liberati:2020mdr}, although the implications for entanglement entropy have not been fully explored.} Fortunately, thanks to excellent previous works, we can analyze the effect of the backreaction on the entanglement entropy analytically and explicitly just by combining the known results appropriately. Since the equation of the background subject to backreaction has been known, we can get its explicit form just by solving it. 
The wave equation in such an analog BH background is also well-known and we can explicitly construct mode functions in the asymptotic region. Furthermore in \cite{Coutant:2017nea}, by focusing on the very low-energy modes, the authors elegantly derived the linear transformation between them (the Bogoliubov transformation), obtaining analytic expressions perturbatively. These results provide a crucial ingredient for the analysis in the present work. On the other hand, a convenient parameterization of the Bogoliubov transformation found in \cite{Nambu:2021lix} enables us to express the entanglement entropy in terms of the Bogoliubov coefficients. Then by studying their changes due to the backreaction, we can concretely estimate the effect of the backreaction on the entanglement entropy. 

The organization of this paper is as follows: in section \ref{sec:theory}, we introduce the basic equations for the background and the fluctuations around it in the BEC. 
We mention its relation to the scalar field on the BH background and to the equations of fluid. We also define our setup of a step-like configuration of the BEC. In section \ref{sec:mode}, we identify the mode functions on such a background. It is found that mode functions available inside and outside the analog BH are distinct due to difference of velocity of the flow. In section \ref{sec:EE}, we review the derivation of the Bogoliubov coefficients in \cite{Coutant:2017nea} and clarify how the entanglement entropy of the analog Hawking radiation can be rewritten in terms of them. We then find an explicit form of backreaction in section \ref{sec:backreaction} and estimate how it affects the entanglement entropy through deformation of the Bogoliubov coefficient. The final section \ref{sec:discussions} is devoted to conclusions and discussions.  

Appendix \ref{app:rhor} provides details of derivation of the reduced density matrix which is necessary to get the entanglement entropy.

\section{Theory of Bose-Einstein condensates}
\label{sec:theory}

In this section we review the basic theory of Bose-Einstein condensate (BEC) and define our setup. 

\subsection{Basic equations}
\label{subsec:basic}
We start from the Hamiltonian of BEC in $d$-spatial dimensions 
\[
H=\int d^dx \left\{\phi^\dagger(x)
\left(-\frac{1}{2m}\bm\nabla^2+U(x)\right)\phi(x)
+\frac12\lambda(x)\left(\phi^\dagger(x)\phi(x)\right)^2\right\},
\label{eq:BEC}
\]
where $\phi(x)$ is a scalar field describing BEC and we have set $\hbar=1$. $\lambda(x)\geq 0$ and $U(x)$ are the coupling ``constant'' and the potential, respectively, but for later discussion we allow the spatial dependence of $\lambda$. By the equal-time commutation relation 
\[
&[\phi(x),\phi^\dagger(y)]=\delta^d(x-y), \nn \\
&[\phi(x),\phi(y)]=0,
\label{eq:CCR}
\]
the Heisenberg equation becomes 
\[
i\partial_t\phi(x)&=[\phi(x),H] \nn \\
&=\left(-\frac{1}{2m}\bm\nabla^2+U(x)\right)\phi(x)
+\lambda(x)\phi^\dagger(x)\phi(x)^2.
\label{eq:Heisenberg}
\]
Taking the vacuum expectation value (VEV) in both sides, we obtain the Gross-Pitaevskii (GP) equation 
for $\phi_0(x)=\vev{\phi(x)}$
\[
i\partial_t\phi_0(x)=\left(-\frac{1}{2m}\bm\nabla^2+U(x)\right)\phi_0(x)+\lambda(x)\phi_0^\dagger(x)\phi_0(x)^2.
\label{eq:GP}
\]
Here we assumed that $\vev{\phi(x)^n}=\vev{\phi(x)}^n$ as in the mean-field approximation. We consider small fluctuation of the original field $\phi(x)$ around the background $\phi_0(x)$ as  
$\phi(x)=\phi_0(x)+\delta\phi(x)$. Plugging this into \eqref{eq:Heisenberg}
yields the Bogoliubov-de Gennes (BdG) equation
\[
i\partial_t\delta\phi(x)
=&\left(-\frac{1}{2m}\bm\nabla^2+U(x)\right)\delta\phi(x)
\nn \\
&+\lambda(x)\left(2\phi_0(x)^\dagger\phi_0(x)\delta\phi(x)
+\phi_0(x)^2\delta\phi(x)^\dagger\right).
\label{eq:BdG}
\]
Here we used the GP equation \eqref{eq:GP}. 
In later subsection we will see that this equation agrees with that of a scalar field in the Schwarzschild black hole (BH) background. Thus $\phi_0$ and $\delta\phi$ are analogs of the black hole and the Hawking radiation, respectively. In general, in the presence of nontrivial $\delta\phi$, the background $\phi_0$ is subject to backreaction. 
In \cite{PJ,Liberati:2020mdr} the equation of the background subject to such backreaction is given by modifying \eqref{eq:GP} to  
\[
i\partial_t\wt{\phi}_0 
=\left(-\frac{1}{2m}\bm\nabla^2+U\right)\wt{\phi}_0
+\lambda(x)\left[\wt{\phi}_0 ^\dagger\wt{\phi}_0 ^2
+\left(
2\wt{\phi}_0 \vev{\delta\phi^\dagger\delta\phi}
+\wt{\phi}_0 ^\dagger\vev{\delta\phi^2}
\right)\right].
\label{eq:backreactedGP}
\]
It is reasonable to regard the backreaction $\wt{\phi}_0-\phi_0$ as higher order than $\delta\phi$. Hence in order to derive this equation, we first expand the original field $\phi$ by introducing a formal expansion parameter $\thbar$ to keep track of orders in the fluctuation as\footnote{A similar argument is provided in \cite{Pal:2024qno}.} 
\[
\phi=\phi_0+\thbar\delta^{(1)}\phi+\thbar^2\delta^{(2)}\phi+\thbar^3\delta^{(3)}\phi
+\cdots. 
\label{eq:exp}
\]
Then substituting this for eq.\eqref{eq:Heisenberg}, we obtain 
eqs.\eqref{eq:GP}, \eqref{eq:BdG} at ${\cal O}(\thbar^0)$, ${\cal O}(\thbar^1)$, respectively with $\delta^{(1)}\phi=\delta\phi$. 
Then adding ${\cal O}(\thbar^2)$ contribution and \eqref{eq:GP}, we find that 
its VEV is given as 
\[
&i\partial_t\left(\phi_0+\thbar^2\vev{\delta^{(2)}\phi}\right) \nn \\
&=\left(-\frac{1}{2m}\bm\nabla^2+U\right)
\left(\phi_0+\thbar^2\vev{\delta^{(2)}\phi}\right)
\nn \\
&\phantom{=}+\lambda\left[\phi_0^\dagger\phi_0^2
+\thbar^2\left(\phi_0^2\vev{\delta^{(2)}\phi^\dagger}
+2\phi_0^\dagger\phi_0\vev{\delta^{(2)}\phi}
+2\phi_0\vev{\delta^{(1)}\phi^\dagger\delta^{(1)}\phi}
+\phi_0^\dagger\vev{\delta^{(1)}\phi^2}
\right)\right].
\label{eq:order2}
\]
In fact, up to ${\cal O}(\thbar^2)$ of our interest, this equation agrees with \eqref{eq:backreactedGP} under identification $\wt{\phi}_0=\phi_0+\thbar^2\vev{\delta^{(2)}\phi}$.

\subsection{Comparison to fluid}
\label{subsec:fluid}
In this subsection we see correspondence between the basic equations of the BEC and of fluid, namely microscopic and macroscopic parameters.  

Setting 
\[
\phi_0=\sqrt{\rho_0}e^{i\theta_0}
\label{eq:polar}
\]
in \eqref{eq:GP}, we find that as usual
\[
&\partial_t\rho_0=-\bm\nabla\cdot\bm J, \nn \\
&\bm J=\frac{1}{2mi}
\left(\phi_0^\dagger\bm\nabla\phi_0-\phi_0\bm\nabla\phi_0^\dagger\right)
=\rho_0\bm v,
\quad\quad \text{where}\qquad 
\bm v\equiv \frac{\bm\nabla\theta_0}{m}.
\label{eq:v} 
\]
$\rho_0$ and $v$ are thus the density and velocity of BEC, respectively. 
We also have 
\[
\partial_t\theta_0=\frac{1}{2m}
\left(\frac{1}{\sqrt{\rho_0}}\bm\nabla^2\sqrt{\rho_0}-\left(\bm\nabla\theta_0\right)^2
\right)-U-\lambda\rho_0.
\label{eq:dtheta}
\]
Comparing this equation with the Euler equation of fluid without a vortex 
\[
\partial_t\bm v=-\bm\nabla\left[\frac12\bm v^2+h(\rho)+\frac{U}{m}\right],
\]
we can identify the enthalpy $h(\rho_0)$ as 
\[
h(\rho_0)=\frac{\lambda}{m}\rho_0,
\]
from which we get the sound velocity $c_s>0$ as 
\[
c_s^2\equiv\frac{dp}{d\rho_0}=\rho_0h'(\rho_0)=\frac{\lambda}{m}\rho_0
\quad \quad \text{i.e.} \quad c_s^2=\frac{\lambda\rho_0}{m}.
\label{eq:sv}
\]
However, it should be noticed that if $\lambda$ or $\rho_0$ has spatial dependence, 
$c_s$ also depends on $x$. 
In the comparison above, we neglected the first term on the right-hand side in \eqref{eq:dtheta}, 
which does not have a classical counterpart. In fact, this term has $\hbar^2$ if we retrieve the Planck constant and is called the quantum pressure term.

\subsection{Comparison to black hole}
\label{subsec:BH}
In this subsection we review that the BdG equation \eqref{eq:BdG} can be recast as that of a scalar field on the Schwarzschild black hole background, e.g. \cite{Kurita:2008fb}.  
In the BdG equation \eqref{eq:BdG}, we set 
\[
\delta\phi=\phi_0\wh{\delta\phi},
\label{eq:hat}
\]
and use \eqref{eq:polar} to get 
\[
i\partial_t\wh{\delta\phi}
=-\frac{1}{m}\left(\frac{1}{2\rho_0}\bn\cdot\left(\rho_0\bn\wh{\delta\phi}\right)
+i\left(\bn\theta_0\right)\cdot\bn\wh{\delta\phi}\right)
+\lambda\rho_0\left(\wh{\delta\phi}+\wh{\delta\phi}^\dagger\right).
\label{eq:fluc_eq}
\]
The imaginary and real parts read 
\[
i\left(\partial_t+\frac1m\left(\bn\theta_0\right)\cdot\bn\right)
\left(\wh{\delta\phi}+\wh{\delta\phi}^\dagger\right)
=-\frac1m\frac{1}{2\rho_0}\bn\cdot\left(\rho_0\bn\left(\wh{\delta\phi}-\wh{\delta\phi}^\dagger\right)\right),
\label{eq:Im}
\]
and
\[
i\left(\partial_t+\frac1m\left(\bn\theta_0\right)\cdot\bn\right)
\left(\wh{\delta\phi}-\wh{\delta\phi}^\dagger\right)
=-\frac1m\frac{1}{2\rho_0}\bn\cdot\left(\rho_0\bn\left(\wh{\delta\phi}+\wh{\delta\phi}^\dagger\right)\right)
+2\lambda\rho_0\left(\wh{\delta\phi}+\wh{\delta\phi}^\dagger\right),
\label{eq:Re}
\]
respectively. Note that these equations are satisfied even if $\phi_0$ is not constant. On the right-hand side in eq.\eqref{eq:Re}, the first term comes from the quantum pressure term and hence can be ignored in the presence of the second term. 
We thus obtain the closed equation for $\wh{\delta\phi}-\wh{\delta\phi}^\dagger$ as 
\[
-\left(\partial_t+\bn\cdot\bm v\right)\frac{\rho_0}{c_s^2}
\left(\partial_t+\bm v\cdot\bn\right)
\left(\wh{\delta\phi}-\wh{\delta\phi}^\dagger\right)
+\bn\cdot\left(\rho_0\bn\left(\wh{\delta\phi}-\wh{\delta\phi}^\dagger\right)\right)
\simeq 0,
\label{eq:closedeq}
\]
where $\simeq$ means ignoring the quantum pressure term and we used an identity for an arbitrary function $f$
\[
\left(\partial_t+\bn\cdot\bm v\right)\rho_0f
=\rho_0\left(\partial_t+\bm v\cdot\bn\right)f.
\label{eq:formula}
\]
Comparing this equation with that of the massless scalar field on a curved background 
\[
\frac{1}{\sqrt{g}}\partial_\mu
\left(\sqrt{-g}g^{\mu\nu}\partial_{\nu}\left(\wh{\phi}-\wh{\phi}^\dagger\right)\right)=0,
\]
we find that the background has essentially the Schwarzschild metric:
\[
ds^2=&\frac{\rho_0}{c_s}
\left(-\left(1-\frac{v_i^2}{c_s^2}\right)\left(c_sd\tau\right)^2
+\frac{1}{1-\frac{v_i^2}{c_s^2}}d{x^i}^2\right)
\label{eq:SchBH}
\]
with
\[
d\tau=dt-\frac{v_i}{c_s^2-v_i^2}dx^i.
\]
We also recognize that $c_s$ plays a role of the velocity of light $c$, 
and hence the horizon is characterized as the sonic point where $|\bm v|=c_s$.  
However, it is again noted that $c_s^2$ is in general $x^i$-dependent in contrast to $c$.

\subsection{Set up}
\label{subsec:setup}
Now let us specify our configuration, namely choice of the background $\phi_0$. 
In the following let us consider a case well-studied in the literature e.g. 
\cite{Macher:2009tw,Macher:2009nz,Fabbri:2010tj,Mayoral:2010un,Coutant:2017nea,Curtis:2018qey,Nambu:2021lix,Osawa:2022vkw}, where $\phi_0$ has nontrivial spatial dependence only in the $x$-direction, and assume that excitations along other spatial directions have much higher energies than those along $x$-direction. In this paper we concentrate on quite low energy excitation and therefore, the system has excitations only along $x$-direction and is effectively in $1+1$-dimensions described by $t,x$. 

We first assume that the background density is static: $\partial_t\rho_0=0$. 
Since we are interested in a set up analogous to the BH background 
with $c$ replaced with $c_s$, we need to construct a configuration 
where the fluid of the condensate flows, e.g. in the negative $x$-direction: $v<0$, and $c_s>-v$ is realized outside the horizon, while inside the horizon, $c_s<-v$ so that the sound cannot go through the horizon. For simplicity, we take the origin of $x$ at the horizon and $v(x)$ is assumed to be piecewise constant as follows: 
\[
v(x)=\left\{
\begin{array}{cc}
   v_r & (x>0) \\
   -c_s & (x=0) \\
   v_l & (x<0)
\end{array}\right.
\quad \text{with} \quad 0<-v_r<c_s<-v_l.
\label{eq:vdef}
\]
One of the most convenient choices would be as in e.g. 
\cite{Macher:2009tw,Macher:2009nz,Nambu:2021lix}
\[
v(x)=c_s\left(-1+D\tanh\left(\frac{x}{\epsilon}\right)\right)\qquad (\epsilon\rightarrow 0).
\label{eq:profile}
\]
Given $v(x)$ above, we obtain $\theta_0(x)$ from \eqref{eq:v} as 
\[
\theta_0(x)=m\int^{x}v\,dx.
\label{eq:theta0}
\]
Eq.\eqref{eq:v} imposes $\rho_0(x)=-\alpha/v(x)$ with $\alpha$ positive constant.\footnote{In the case of eq.\eqref{eq:profile}, $\rho_0\geq0$ requires $D\leq 1$.} In particular, we notice that if we consider a static background and a non-constant velocity as in eq.\eqref{eq:vdef}, the condensate cannot be homogenous. Furthermore, in this case eq.\eqref{eq:sv} implies that $c_s$ cannot be also constant in $x$.  For technical reason discussed later, we restrict ourselves to the case with constant $c_s$, and hence $\lambda$ must have $x$-dependence as $\lambda(x)=mc_s^2/\rho_0(x)=-mc_s^2v(x)/\alpha$. However, as in eq.\eqref{eq:vdef} $\lambda(x)$ is at least piecewise constant. 
We thus take the concrete form of background of condensate as the plane wave 
\[
\phi_0(x,t)&=e^{-i\mu t}\bar{\phi}_0(x), \nn \\
\bar{\phi}_0(x)&=\sqrt{\rho_0(x)}e^{i\theta_0(x)}.
\label{eq:back}
\]
Here $\theta_0(x)$ is given by eq.\eqref{eq:theta0} with eq.\eqref{eq:vdef}. 
Since $\rho_0(x)$ and $v(x)$ are piecewise constant, straightforward calculation shows that 
this plane wave indeed satisfies the GP equation provided
\[
\mu=\frac12 mv^2+U+\lambda\rho_0.
\label{eq:mu}
\]
However, we note that it is no longer a solution at the origin where $v(x)$ has discontinuity. 
In order to find its precise form, we need to regularize the configuration as in \eqref{eq:profile} 
and to solve the GP equation directly. In later section we show that we do not have to do this for the purpose of calculating the entanglement entropy from the Bogoliubov coefficients, which are given in terms of relations between modes in the asymptotic regions.\footnote{It should however be kept in mind that $\mu$ is different between $x>0$ and $x<0$. As we will see later, the equations for the fluctuation $\wh{\delta\phi}$ do not depend on $\mu$, and hence this fact does not cause any apparent problem at least for the Bogoliubov coefficients and the entanglement entropy. It may be also possible to bypass this by considering a case with a constant $v$ and $x$-dependent $c_s$.}

\subsection{Effect of backreaction on radiation}
\label{subsec:back-reactedBdG}
Since we are interested in how backreaction affects the entanglement entropy 
of the analog Hawking radiation, we need an equation for fluctuation around the background subject to the backreaction taken into account in eq.\eqref{eq:backreactedGP}. 
When we derive the BdG equation, we expand the GP equation around its solution. Thus we naively expect that we just repeat the same procedure for the GP equation with the backreaction in eq.\eqref{eq:backreactedGP}. 
However, if we do this, it is not clear which fluctuation belongs to 
the background or the radiation, because eq.\eqref{eq:backreactedGP} itself includes the two-point function of the fluctuation. In order to clarify this, we continue the expansion 
\eqref{eq:exp} further. Plugging eq.\eqref{eq:exp} into the original Heisenberg equation 
\eqref{eq:Heisenberg}, at ${\cal O}(\thbar^3)$ we have 
\[
&i\partial_t\delta^{(3)}\phi
=\left(-\frac{1}{2m}\bm\nabla^2+U\right)\delta^{(3)}\phi
\nn \\
&+\lambda\left(\phi_0^2\delta^{(3)}\phi^\dagger
+2\delta^{(1)}\phi
\left(\phi_0\delta^{(2)}\phi^\dagger+\phi_0^\dagger\delta^{(2)}\phi\right)
+2\delta^{(1)}\phi^\dagger
\left(\delta^{(1)}\phi^2+2\phi_0\delta^{(2)}\phi\right)
+2\phi_0^\dagger\phi_0\delta^{(3)}\phi
\right).
\]
Adding this equation and that at ${\cal O}(\thbar^1)$ gives rise to 
\[
&i\partial_t\left(\delta^{(1)}\phi+\thbar^2\delta^{(3)}\phi\right)
=\left(-\frac{1}{2m}\bm\nabla^2+U\right)
\left(\delta^{(1)}\phi+\thbar^2\delta^{(3)}\phi\right)
\nn \\
&+\lambda\biggl\{
\left(\delta^{(1)}\phi+\thbar^2\delta^{(3)}\phi\right)^\dagger
\left(\phi_0+\thbar^2\delta^{(2)}\phi\right)^2
+2\left(\phi_0+\thbar^2\delta^{(2)}\phi\right)^\dagger
\left(\phi_0+\thbar^2\delta^{(2)}\phi\right)
\left(\delta^{(1)}\phi+\thbar^2\delta^{(3)}\phi\right) \nn \\
&\phantom{+\lambda\Biggl\{}+\thbar^2
\left(\delta^{(1)}\phi+\thbar^2\delta^{(3)}\phi\right)^\dagger
\left(\delta^{(1)}\phi+\thbar^2\delta^{(3)}\phi\right)^2
\biggr\}\Biggr|_{\text{up to }{\cal O}(\thbar^2)}.
\]
Namely, if we regard the backreacted background as $\wt{\phi}_0=\phi_0+\thbar^2\delta^{(2)}\phi$, and the fluctuation around it as $\delta\wt{\phi}=\delta^{(1)}\phi+\thbar^2\delta^{(3)}\phi$, then we obtain 
\[
i\partial_t\delta\wt{\phi}
=&\left(-\frac{1}{2m}\bm\nabla^2+U\right)\delta\wt{\phi}
+\lambda\biggl\{\delta\wt{\phi}^\dagger\wt{\phi}_0^2
+2\wt{\phi}_0^\dagger\wt{\phi}_0\delta\wt{\phi}
+\thbar^2\delta\wt{\phi}^\dagger\delta\wt{\phi}^2
\biggr\}\Biggr|_{\text{up to }{\cal O}(\thbar^2)}. 
\label{eq:order3}
\]
This equation is regarded as the linearized one for the fluctuation $\delta\wt\phi$. We therefore  
replace its coefficients with their VEV because the differences are of higher order. 
We thus obtain 
\[
i\partial_t\delta\wt{\phi}
=\left(-\frac{1}{2m}\bm\nabla^2+U\right)\delta\wt{\phi} 
+\lambda\left(
2\left(\left|\wt{\phi}_0\right|^2+\thbar^2\vev{\delta\wt{\phi}^\dagger\delta\wt{\phi}}\right)
\delta\wt{\phi}
+\left(\wt{\phi}_0^2+\thbar^2\vev{\delta\wt{\phi}^2}\right)\delta\wt{\phi}^\dagger
\right).
\label{eq:brBdGfluc}
\]
This result with $\thbar=1$ indeed reproduces the naive expansion of the backreacted GP equation \eqref{eq:backreactedGP} mentioned above. However, it is now clear which fluctuation belongs to the backreaction or the radiation. Moreover, setting as in \eqref{eq:hat} 
\[
\delta\wt{\phi}=\wt{\phi}_0\wh{\delta\wt{\phi}},
\]
and using \eqref{eq:backreactedGP}, we have 
\[
i\wt{\phi}_0\partial_t\wh{\delta\wt{\phi}} 
=
-\frac{1}{m}\bm\nabla\wt{\phi}_0\bm\nabla\wh{\delta\wt{\phi}}
-\frac{1}{2m}\wt{\phi}_0\bm\nabla^2\wh{\delta\wt{\phi}} 
+\lambda\wt{\phi}_0^\dagger\wt{\phi}_0^2
\left(\wh{\delta\wt{\phi}}+\wh{\delta\wt{\phi}}^\dagger\right). 
\]
It agrees with eq.\eqref{eq:fluc_eq} with $\phi_0$ replaced with $\wt{\phi}_0$. 
Namely, the equation for the fluctuation $\wh{\delta\wt{\phi}}$ around the backreacted background is obtained by making replacement 
of the background $\phi_0\rightarrow\wt{\phi}_0$ from that without the backreaction. 
It means that for the radiation, the effects of the backreaction are all encoded 
into this replacement, and that no other new term arises in its equation. It comes from the fact that the background and the radiation are definitely separated as above 
and that even the backreaction behaves as a part of the background in the mean-field theory and does not have a nontrivial correlation function with the fluctuation.

\subsection{Comparison to genuine Hawking radiation}
\label{subsec:comparison}
In this subsection we make a comparison to genuine BH physics. 
We first note that in the BEC, 
the original field $\phi$ is decomposed perturbatively as in \eqref{eq:exp} 
and such hierarchical structure enables us to give the physically clear interpretation to each term as background, analog Hawking radiation, and backreaction.  
In particular, we notice that the backreaction and its source (the two-point functions) are of ${\cal O}(\thbar^2)$.\footnote{Its explicit form is given in eq.\eqref{eq:backreactionform}.} 
Later we will show that the former plays a role of decreasing the entanglement entropy, which is originally of ${\cal O}(\thbar^0)$ in the sense that it is given by the parameter of the background, more precisely, the Bogoliubov coefficients. Thus the correction to the entanglement entropy due to the backreaction starts at ${\cal O}(\thbar^2)$ and there are no ${\cal O}(\thbar^1)$ corrections. This result is attributed to the fact that from our construction, correlation function between $\phi_0$ and $\delta^{(1)}\phi$ vanishes as $\vev{\phi_0\delta^{(1)}\phi_0}=\phi_0\vev{\delta^{(1)}\phi}=0$, which would be of ${\cal O}(\thbar^1)$ if it were nonzero, and hence in \eqref{eq:order2} the backreaction contributes at ${\cal O}(\thbar^2)$.  

Let us compare these observations to the BH case where we have the Einstein equation 
\[
R_{\mu\nu}-\frac12Rg_{\mu\nu}=8\pi G\vev{T_{\mu\nu}}. 
\label{eq:Einstein}
\]
Our strategy is similar to the procedure to solve this equation perturbatively 
\cite{Balbinot:2025upx}. Namely, we first solve it in absence of the right-hand side to get the BH solution. Next in this background we consider the matter field and evaluate $T_{\mu\nu}$. Then we find the backreaction to solve \eqref{eq:Einstein} again in the presence of the right-hand side. They correspond to the GP equation, the BdG equation, and \eqref{eq:backreactedGP}, respectively. However, as we have seen 
in subsection \ref{subsec:BH}, direct correspondence is quite limited: we only have the explicit relation between the equation of motion for the scalar field on the BH background and the BdG equation. In particular, the existence of the backreaction implies that of the energy-momentum tensor and it is desirable to find a direct correspondence between \eqref{eq:backreactedGP} and \eqref{eq:Einstein}. 
However, at least we can observe several similarities:  
we also expect that the correlation function between the BH and the Hawking radiation 
is quite suppressed but nonvanishing in contrast to the BEC case, since both carry energy and necessarily interact gravitationally. The source of the backreaction of the quadratic order also resembles the energy-momentum tensor which is generally quadratic in the matter field or quantum fluctuations. 
On the other hand, one of the biggest differences between the BEC and the BH is that in the latter the BH and the Hawking radiation do not appear as different orders in expansion of a single field as in \eqref{eq:exp}. However, in nonperturbative formulation of string theory based on a matrix model \cite{Banks:1996vh,Ishibashi:1996xs}, we expect that the space-time, gauge field and matter are unified in a matrix as the diagonal and the off-diagonal elements. From a speculative viewpoint, we hope studies in this paper would help in description of the Hawking radiation in this kind of matrix models.

\section{Mode functions}
\label{sec:mode}

In this section we analyze properties of fluctuation $\delta\phi$ on our background $\phi_0$ specified in eq.\eqref{eq:back}. For this purpose, we begin by reviewing a general argument on the fluctuation in $d+1$-dimensions. 

\subsection{Bogoliubov equation}
\label{subsec:Bogoliubov}
We first separate the time dependence of the background as in \eqref{eq:back}
\[
\delta\phi=e^{-i\mu t}{\delta\bar{\phi}}.
\label{eq:bdeltaphi}
\]
The BdG equation \eqref{eq:BdG} becomes 
\[
i\partial_t{\delta\bar{\phi}}=&{\cal L}{\delta\bar{\phi}}+\lambda\bar{\phi}_0^2{\delta\bar{\phi}}^{\dagger},
\nn \\
&{\cal L}\equiv -\frac1{2m}\bm\nabla^2+U-\mu+2\lambda\left|\bar{\phi}_0\right|^2. 
\label{eq:BdGonbackground}
\]
Suppose that we have the mode functions $u_k(x)$, $v_k(x)$ labeled by the $d$-dimensional momentum $k$ and that we make an expansion in terms of them as 
\[
\delta\bar{\phi}(x,t)=\int\frac{d^dk}{(2\pi)^d}
\left(u_{k}(x)e^{-i\epsilon_kt}a_k-v_k^*(x)e^{i\epsilon_kt}a_k^{\dagger}\right),
\label{eq:modeexp}
\]
then a straightforward calculation shows that the BdG equation is equivalent to the Bogoliubov equation 
\[
\left\{
\begin{array}{ll}
&{\cal L}u_k(x)-\lambda\bar{\phi}_0^2v_k(x)=\epsilon_ku_k(x), \\
&{\cal L}v_k(x)-\lambda\bar{\phi}_0^{\dagger2}u_k(x)=-\epsilon_kv_k(x).
\end{array}\right. 
\label{eq:Bogoliubov}
\]
We impose the standard commutation relation
\[
[a_k,a_{k'}^{\dagger}]=(2\pi)^d\delta^d(k-k'),
\]
and from eq.\eqref{eq:CCR} we obtain 
\[
&[\delta\bar\phi(x),\delta\bar{\phi}^{\dagger}(y)]=\delta^d(x-y), \nn \\
&[\delta\bar\phi(x),\delta\bar{\phi}(y)]=0.
\]
Then it is easy to find that $u_k$ and $v_k$ should satisfy the following relation 
\[
&\int\frac{d^dk}{(2\pi)^d}\left(u_k(x)u_k^*(y)-v_k^*(x)v_k(y)\right)=\delta^d(x-y), \nn \\
&\int\frac{d^dk}{(2\pi)^d}\left(u_k(x)v_k^*(y)-v_k^*(x)u_k(y)\right)=0.
\label{eq:uvrelation1}
\]
Using these relations, we can invert eq.\eqref{eq:modeexp} as
\[
a_k=\int d^dx\left(u_k^*(x)\delta\bar{\phi}(x)+v_k^*(x)\delta\bar{\phi}^{\dagger}(x)\right),
\]
which gives  
\[
&\int d^dx\left(u_k^*(x)u_{k'}(x)-v_k^*(x)v_{k'}(x)\right)=(2\pi)^d\delta^d(k-k'), \nn \\
&\int d^dx\left(u_k(x)v_{k'}(x)-v_k(x)u_{k'}(x)\right)=0.
\label{eq:uvrelation2}
\]
It is well-known that the above modes diagonalize the Hamiltonian for the fluctuation\footnote{This Hamiltonian is understood as constructed in such a way that will reproduce  \eqref{eq:BdGonbackground}, or obtained directly from the original BEC Hamiltonian \eqref{eq:BEC} up to the quadratic order in $\delta\bar{\phi}$ with noting the time-dependent 
canonical transformation.}  
\[
H_2=\int d^dx \left\{\delta\bar{\phi}^\dagger
{\cal L}\delta\bar{\phi}
+\frac12\lambda(x)\left(\bar{\phi}_0^2\delta\bar{\phi}^{\dagger2}+\bar{\phi}_0^{\dagger2}\delta\bar{\phi}^2\right)\right\},
\label{eq:Hforfluc}
\]
as 
\[
H_2=\int\frac{d^dk}{(2\pi)^d}\epsilon_k\left(a_k^{\dagger}a_k-\int d^dx|v_k(x)|^2\right).
\]

In what follows let us consider the $U\equiv 0$ case. In order to solve the Bogoliubov equation \eqref{eq:Bogoliubov}, we make an ansatz 
\[
u_k(x)&=e^{i(mv+k)x}U_k, \nn \\
v_k(x)&=e^{-i(mv-k)x}V_k,
\label{eq:ansatz}
\]
then by using \eqref{eq:mu}, in the region where $v$ is constant, 
eq.\eqref{eq:Bogoliubov} becomes 
\[
\begin{pmatrix}
\frac{k^2}{2m}+vk+\lambda\rho_0 & -\lambda\rho_0 \\
\lambda\rho_0 & -\frac{k^2}{2m}+vk-\lambda\rho_0
\end{pmatrix}
\begin{pmatrix}
U_k \\
V_k
\end{pmatrix}
=\epsilon_k
\begin{pmatrix}
U_k \\
V_k
\end{pmatrix}.
\]
Hence a nontrivial solution exists only when the dispersion relation 
\[
\epsilon_k=vk\pm\sqrt{\epsilon_k^{(0)}(\epsilon_k^{(0)}+2\lambda\rho_0)}
\label{eq:dispersiontemp}
\]
is satisfied, where $\epsilon_k^{(0)}\equiv k^2/(2m)$. 
Furthermore, eq.\eqref{eq:uvrelation2} implies that if we require $U_k,V_k\in\bm R$, 
we have to take $+$ sign in the above equation: 
\[
\epsilon_k=vk+\sqrt{\epsilon_k^{(0)}(\epsilon_k^{(0)}+2\lambda\rho_0)},
\label{eq:dispersion}
\]
and that 
\[
u_k(x)&=e^{i(mv+k)x}\frac{\sqrt{\epsilon^{(0)}_k+2\lambda\rho_0}+\sqrt{\epsilon^{(0)}_k}}{2\left(\epsilon^{(0)}_k\left(\epsilon^{(0)}_k+2\lambda\rho_0\right)\right)^{\frac14}}, \nn \\
v_k(x)&=e^{-i(mv-k)x}\frac{\sqrt{\epsilon^{(0)}_k+2\lambda\rho_0}-\sqrt{\epsilon^{(0)}_k}}{2\left(\epsilon^{(0)}_k\left(\epsilon^{(0)}_k+2\lambda\rho_0\right)\right)^{\frac14}},
\label{eq:uvresult}
\]
which also satisfies eq.\eqref{eq:uvrelation1}. It should be noted that the above equation is derived under the assumption that $\lambda\rho_0$ is constant. In fact, if $\lambda\rho_0$ is not constant, the above result is no longer true. For example, when we have non-homogeneous condensate with constant $\lambda$, the above equation cannot satisfy \eqref{eq:uvrelation2} in general. This is one of the reasons why we allow $x$-dependence of $\lambda$.

\subsection{Identification of momenta}
\label{subsec:momenta}
In this subsection we review the results in identification of momenta satisfying \eqref{eq:dispersion} which label eq.\eqref{eq:uvresult} in the case of one spatial dimension. 

In order to find momenta in \eqref{eq:dispersion} given $\epsilon_k=\omega$, it is convenient 
to consider intersections between the straight line $\omega-vk$ and the curve $\sqrt{\epsilon_k^{(0)}(\epsilon_k^{(0)}+2\lambda\rho_0)}$ as functions of the one-dimensional momentum $k$ according to eq.\eqref{eq:dispersion}. Let us take a closer look at them 
by considering positive or negative $x$ separately. In computation of the Bogoliubov coefficient, 
we need to specify which mode is incoming or outgoing from the horizon point of view, 
by examining its group velocity. Namely, in the right region ($x>0$), a mode with positive (negative) $d\omega /dk$ is identified with the outgoing (ingoing) mode, respectively, and vice versa in the left region ($x<0$).

\begin{enumerate}
\item $x>0$ \\
In the right region, $0<-v_r<c_s$ and as shown in Figure \ref{fig:dispersionright}, the straight line and the curve intersect only at two points 
for an arbitrary $\omega>0$. It is easy to read their group velocity by comparing two slopes at the intersections. Since they have positive $\omega$, the modes \eqref{eq:uvresult} with positive (negative) $k$ are the right- (left-) moving ones, respectively in the sense of the phase velocity. Thus we have two modes as solutions to eq.\eqref{eq:dispersion}, one of which is an outgoing mode denoted by $k_R^{\text{out}}$, while the other is an ingoing one by $k_L^{\text{in}}$. Here $R$ and $L$ indicate the positive and negative group velocity, respectively. Thus these labels should not be confused with the right- and left-moving associated with the phase velocity mentioned above. In later discussions we only use the $R$ and $L$ labels since they determine ingoing- and outgoing modes from the viewpoint of the horizon. More precisely, as shown in \cite{Macher:2009nz}, the mode expansion \eqref{eq:modeexp} can be rewritten in the present case as 
\[
\delta\bar{\phi}(x,t)=&\int_{-\infty}^{\infty}\frac{dk}{2\pi}
\left(u_{k}(x)e^{-i\epsilon_kt}a_k-v_k^*(x)e^{i\epsilon_kt}a_k^{\dagger}\right) \nn \\
=&\left(\int_{0}^{\infty}\frac{dk}{2\pi}+\int_{-\infty}^{0}\frac{dk}{2\pi}\right)
\left(u_{k}(x)e^{-i\epsilon_kt}a_k-v_k^*(x)e^{i\epsilon_kt}a_k^{\dagger}\right) \nn \\
=&\int_{0}^{\infty}\frac{d\omega}{2\pi}
\left(e^{-i\omega t}u_{k_{R}^{\text{out}}(\omega)}(x)a_{k_{R}^{\text{out}}(\omega)}
     -e^{i\omega t}v_{k_{R}^{\text{out}}(\omega)}^*(x)a_{k_{R}^{\text{out}}(\omega)}^{\dagger}
\right) \nn \\
+&\int_{0}^{\infty}\frac{d\omega}{2\pi}
\left(e^{-i\omega t}u_{k_{L}^{\text{in}}(\omega)}(x)a_{k_{L}^{\text{in}}(\omega)}
     -e^{i\omega t}v_{k_{L}^{\text{in}}(\omega)}^*(x)a_{k_{L}^{\text{in}}(\omega)}^{\dagger}  
\right).
\label{eq:modeexpright}
\]
Here in the last expression the first and second term are contributions from the right- ($k>0$) and left-moving ($k<0$) modes, respectively. As we can see from Figure \ref{fig:dispersionright}, the $R$, $L$ labels and the right-, left-moving agree with each other, respectively in $x<0$ case.

\begin{figure}[H]
\centering
\centering \includegraphics[height=2.2in]{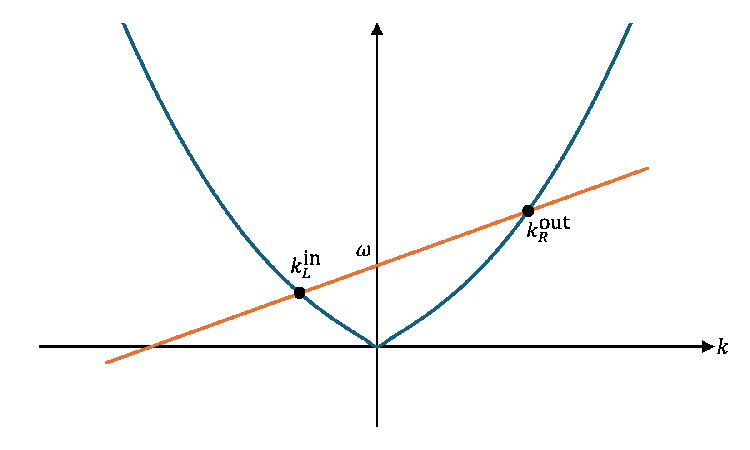}
\vspace{-0cm}
\caption{In $x>0$ region for $\omega>0$, $\omega-vk$ and $\sqrt{\epsilon_k^{(0)}(\epsilon_k^{(0)}+2\lambda\rho_0)}$ are depicted as functions of the momentum $k$. The two intersections are the solution to \eqref{eq:dispersion}. One of them is an outgoing mode with a positive group velocity, and the other is an ingoing one with a negative group velocity.}
\label{fig:dispersionright}
\end{figure}

\item $x<0$ \\
In the left region, $0<c_s<-v_l$ and the straight line and the curve again intersect only at two points for arbitrary $\omega>0$ as shown in Figure \ref{fig:dispersionleft0}. 

\begin{figure}[H]
\centering
\centering \includegraphics[height=2.2in]{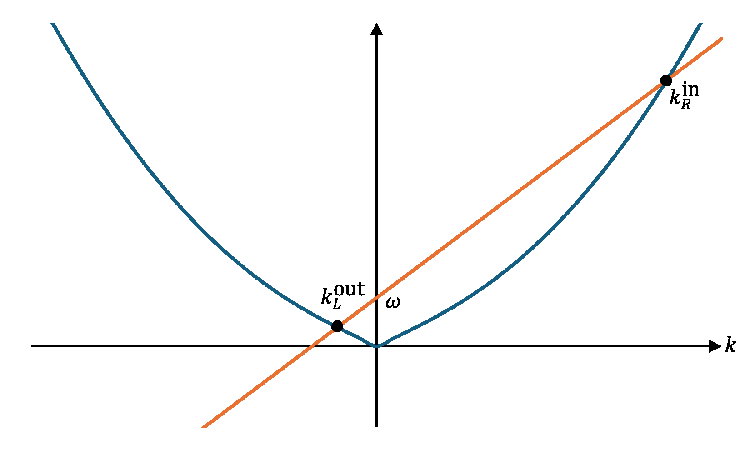}
\vspace{-0cm}
\caption{In $x<0$ region for $\omega>0$, we again have two modes. The right-moving ($k>0$) mode is an ingoing one with a positive group velocity while the left-moving ($k<0$) mode is outgoing with a negative group velocity.}
\label{fig:dispersionleft0}
\end{figure}

However, in this case, we also have two intersections 
for negative $\omega$ satisfying $-\omega_c<\omega<0$ as shown in Figure \ref{fig:dispersionleft}, 
where 
\[
-\omega_c=k_cv+\sqrt{\epsilon^{(0)}_{k_c}\left(\epsilon^{(0)}_{k_c}+2\lambda\rho_0\right)}<0,
\]
with
\[
k_c^2=\frac12m^2\left(v_l^2-4c_s^2+v_l\sqrt{v_l^2+8c_s^2}\right).
\]
Thus even in this case we have two modes as solutions to eq.\eqref{eq:dispersion}, one of which is an ingoing mode denoted by $k_{\bar R}^{\text{in}}$, while the other is an outgoing one by $k_{\bar L}^{\text{out}}$. As we will see in a moment, since they correspond to modes of antiparticles, we put the bar on $R$ and $L$.

\begin{figure}[H]
\centering
\centering \includegraphics[height=2.2in]{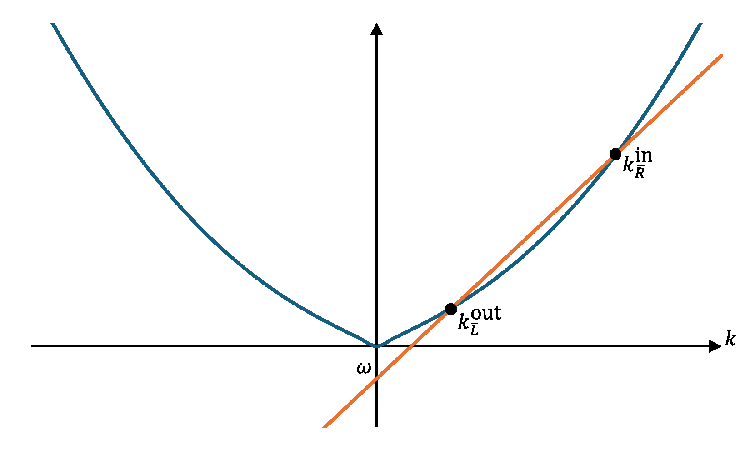}
\vspace{-0cm}
\caption{In $x<0$ region for $-\omega_c<\omega<0$, we still have two intersections. They are the right-moving modes ($k>0$). One of them is an ingoing mode with a positive group velocity, and the other is an outgoing one with a negative group velocity. They are associated with antiparticle excitations.}
\label{fig:dispersionleft}
\end{figure}

Similarly to \eqref{eq:modeexpright}, we write the mode expansion 
\eqref{eq:modeexp} by using the above result. Noting that $\omega=0\leftrightarrow k=2m\sqrt{v_l^2-c_s^2}\equiv k'_c\geq 0$, we obtain  
\[
&\left(\int_{k'_c}^{\infty}\frac{dk}{2\pi}+\int_{-\infty}^{0}\frac{dk}{2\pi}\right)
\left(u_{k}(x)e^{-i\epsilon_kt}a_k-v_k^*(x)e^{i\epsilon_kt}a_k^{\dagger}\right) \nn \\
&=\int_{0}^{\infty}\frac{d\omega}{2\pi}
\left(e^{-i\omega t}\left(u_{k_{R}^{\text{in}}(\omega)}(x)a_{k_{R}^{\text{in}}(\omega)}
+u_{k_{L}^{\text{out}}(\omega)}(x)a_{k_{L}^{\text{out}}(\omega)}\right)\right. \nn \\
&\phantom{=\int_{0}^{\infty}\frac{d\omega}{2\pi}}\left.     
-e^{i\omega t}\left(v_{k_{R}^{\text{in}}(\omega)}^*(x)a_{k_{R}^{\text{in}}(\omega)}^{\dagger}
+v_{k_{L}^{\text{out}}(\omega)}^*(x)a_{k_{L}^{\text{out}}(\omega)}^{\dagger}\right)
\right),
\label{eq:ktoomega1}
\]
and 
\[
&\int_{0}^{k_c}\frac{dk}{2\pi}
\left(u_{k}(x)e^{-i\epsilon_kt}a_k-v_k^*(x)e^{i\epsilon_kt}a_k^{\dagger}\right) \nn \\
&=\int_{-\omega_c}^{0}\frac{d\omega}{2\pi}
\left(
e^{-i\omega t}u_{k_{\bar L}^{\text{out}}(\omega)}(x)a_{k_{\bar L}^{\text{out}}(\omega)}
-e^{i\omega t}v_{k_{\bar L}^{\text{out}}(\omega)}^*(x)a_{k_{\bar L}^{\text{out}}(\omega)}^{\dagger}
\right) \nn \\
&=\int_{0}^{\omega_c}\frac{d\omega}{2\pi}
\left(
e^{-i\omega t}\left(-v_{k_{\bar L}^{\text{out}}(-\omega)}^*(x)a_{k_{\bar L}^{\text{out}}(-\omega)}^{\dagger}\right)
+e^{i\omega t}u_{k_{\bar L}^{\text{out}}(-\omega)}(x)a_{k_{\bar L}^{\text{out}}(-\omega)}\right), 
\label{eq:ktoomega2}
\]
where it should be noticed that in the last line by changing the integration variable 
$\omega\rightarrow -\omega$, the positive and negative frequency modes are interchanged. 
As we will see later, this is the origin where the creation and annihilation operators are mixed 
in the Bogoliubov transformation in the present case. Likewise, 
\[
&\int_{k_c}^{k'_c}\frac{dk}{2\pi}
\left(u_{k}(x)e^{-i\epsilon_kt}a_k-v_k^*(x)e^{i\epsilon_kt}a_k^{\dagger}\right) \nn \\
&=\int_{0}^{\omega_c}\frac{d\omega}{2\pi}
\left(
e^{-i\omega t}\left(-v_{k_{\bar R}^{\text{in}}(-\omega)}^*(x)a_{k_{\bar R}^{\text{in}}(-\omega)}^{\dagger}\right)
+e^{i\omega t}u_{k_{\bar R}^{\text{in}}(-\omega)}(x)a_{k_{\bar R}^{\text{in}}(-\omega)}\right). 
\label{eq:ktoomega3}
\]
Combining eqs.\eqref{eq:ktoomega1}-\eqref{eq:ktoomega3}, we find that \eqref{eq:modeexp} 
now becomes 
\[
\delta\bar{\phi}(x,t)=&\int_{0}^{\omega_c}\frac{d\omega}{2\pi}
\Biggl(
e^{-i\omega t}\biggl(
  u_{k_{R}^{\text{in}}(\omega)}(x)a_{k_{R}^{\text{in}}(\omega)}
+u_{k_{L}^{\text{out}}(\omega)}(x)a_{k_{L}^{\text{out}}(\omega)} \nn \\
&\phantom{\int_{0}^{\omega_c}\frac{d\omega}{2\pi}
\Bigl(e^{-i\omega t}}
-v_{k_{\bar L}^{\text{out}}(-\omega)}^*(x)a_{k_{\bar L}^{\text{out}}(-\omega)}^{\dagger}
-v_{k_{\bar R}^{\text{in}}(-\omega)}^*(x)a_{k_{\bar R}^{\text{in}}(-\omega)}^{\dagger}\biggl) \nn \\
&\phantom{\int_{0}^{\omega_c}\frac{d\omega}{2\pi}}
-e^{i\omega t}\biggl(
  v_{k_{R}^{\text{in}}(\omega)}^*(x)a_{k_{R}^{\text{in}}(\omega)}^{\dagger}
+v_{k_{L}^{\text{out}}(\omega)}^*(x)a_{k_{L}^{\text{out}}(\omega)}^{\dagger} \nn \\
&\phantom{\int_{0}^{\omega_c}\frac{d\omega}{2\pi}\Bigl(e^{-i\omega t}}
-u_{k_{\bar L}^{\text{out}}(-\omega)}(x)a_{k_{\bar L}^{\text{out}}(-\omega)}
-u_{k_{\bar R}^{\text{in}}(-\omega)}(x)a_{k_{\bar R}^{\text{in}}(-\omega)}\biggl) 
\Biggr) 
\label{eq:ktoomegalowE}
\\
+&\int_{\omega_c}^{\infty}\frac{d\omega}{2\pi}
\Biggl(
e^{-i\omega t}\biggl(
  u_{k_{R}^{\text{in}}(\omega)}(x)a_{k_{R}^{\text{in}}(\omega)}
+u_{k_{L}^{\text{out}}(\omega)}(x)a_{k_{L}^{\text{out}}(\omega)}\biggr) \nn \\
&\phantom{\int_{0}^{\omega_c}\frac{d\omega}{2\pi}}
-e^{i\omega t}\biggl(
  v_{k_{R}^{\text{in}}(\omega)}^*(x)a_{k_{R}^{\text{in}}(\omega)}^{\dagger}
+v_{k_{L}^{\text{out}}(\omega)}^*(x)a_{k_{L}^{\text{out}}(\omega)}^{\dagger}\biggr) 
\Biggr). 
\]
\label{eq:ktoomegafull}
\end{enumerate}

\begin{figure}[H]
\centering
\centering 
\includegraphics[clip]{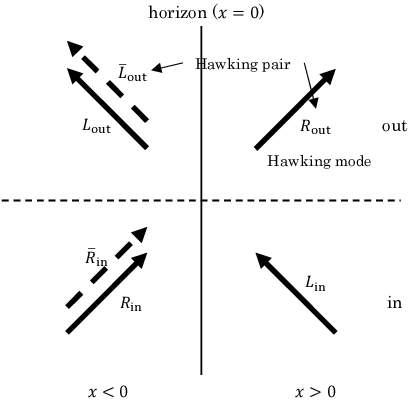}
\vspace{-0cm}
\caption{All modes are classified by the regions and the in/out modes. The antiparticle modes are denoted by dashed lines. The $R_{\text{out}}$ modes are interpreted as the Hawking modes and their Hawking pairs are $\bar{L}_{\text{out}}$ ones.}
\label{fig:modesummary}
\end{figure}

Since we have classified the momentum for fixed $\omega$ completely, we can provide 
an explicit form of each $k$ at least iteratively. For the purpose of analyzing the Hawking radiation, we are interested in the low energy modes. Thus we give an expression of each momentum as a power series with respect to $\omega/(mc_s^2)$ as 
\[
\underline{x>0}\phantom{k_R^{\text{out}}=}& \nn \\
k_R^{\text{out}}=&\frac{\omega}{v_r+c_s}
\biggl(1+{\cal O}\left(\param\right)^2\biggr)>0, \nn \\
k_L^{\text{in}}=&\frac{\omega}{v_r-c_s}
\biggl(1+{\cal O}\left(\param\right)^2\biggr)<0, \nn \\
\underline{x<0}\phantom{k_R^{\text{out}}=}& \nn \\
k_{R}^{\text{in}}=&\bigg(-\frac{\omega v_l}{v_l^2-c_s^2}+2m\sqrt{v_l^2-c_s^2}\biggr)
\biggl(1+{\cal O}\left(\param\right)^2\biggr)>0, \nn \\
k_{\bar R}^{\text{in}}=&\bigg(\frac{\omega v_l}{v_l^2-c_s^2}+2m\sqrt{v_l^2-c_s^2}\biggr)
\biggl(1+{\cal O}\left(\param\right)^2\biggr)>0, \nn \\
k_{L}^{\text{out}}=&\frac{\omega}{v_l-c_s}
\biggl(1+{\cal O}\left(\param\right)^2\biggr)<0, \nn \\
k_{\bar L}^{\text{out}}=&\frac{\omega}{v_l+c_s}
\biggl(1+{\cal O}\left(\param\right)^2\biggr)>0. 
\label{eq:momenta}
\]
Note that in \eqref{eq:ktoomegalowE} 
$k_{\bar R}^{\text{in}}$ and $k_{\bar L}^{\text{out}}$ are subject to the replacement  
$\omega\rightarrow -\omega$. 
These modes are summarized in Figure \ref{fig:modesummary} and the $R_{\text{out}}$ modes are interpreted as the Hawking modes with the Hawking pair $\bar{L}_{\text{out}}$ ones. 

\begin{figure}[H]
\centering
\centering \includegraphics[height=2.2in]{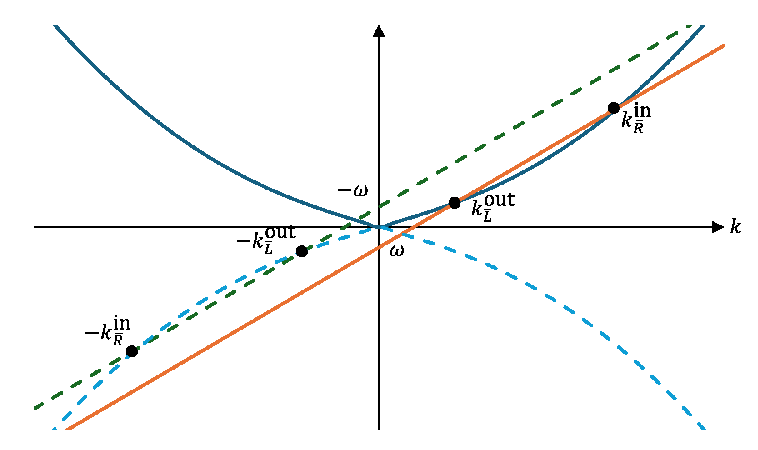}
\vspace{-0cm}
\caption{In $x<0$ region for $\omega<0$, momenta associated with antiparticle excitations are also identified by considering the $-$ sign in \eqref{eq:dispersiontemp} and $\omega\rightarrow -\omega$.}
\label{fig:dispersionleft2}
\end{figure}

Before closing this section, we briefly mention the difference in the treatment of antiparticle modes 
from the literature. We have chosen the $+$ sign in eq.\eqref{eq:dispersiontemp} due to the normalization condition eq.\eqref{eq:uvrelation2}. On the other hand, as in the literature, the momenta $k_{\bar R}^{\text{in}}$, $k_{\bar L}^{\text{out}}$ associated with antiparticle excitations can be also deduced 
by considering the $-$ sign and changing $\omega\rightarrow -\omega$ as shown in Figure \ref{fig:dispersionleft2}. This identification is natural from the point of view in the derivation in \eqref{eq:ktoomega2}.

\section{Entanglement entropy of Hawking radiation}
\label{sec:EE}

In this section we derive the Bogoliubov transformation between modes identified in the previous section, and based on it, we compute the entanglement entropy of the analog Hawking radiation without taking backreaction into account. We will see that as the rate of Hawking pairs increases, it grows as expected. 

\subsection{Bogoliubov transformation}
\label{subsec:Bogoliubov_transf}
In order to analyze the Hawking radiation, we have to find the Bogoliubov transformation between asymptotic ingoing and outgoing modes. 
In the present case, $v(x)$ in \eqref{eq:vdef} is constant in the asymptotic region 
and the modes in the previous section are valid there.  
It is clear that the analog Hawking radiation comes from the modes with $0<\omega<\omega_c$. 
{}From eqs.\eqref{eq:modeexpright} and \eqref{eq:ktoomegalowE}, in the right ($x>0$) region we have one outgoing mode $k_R^{\text{out}}$ and one ingoing mode $k_L^{\text{in}}$, while in the left ($x<0$) region two ingoing modes $k_R^{\text{in}}$, $k_{\bar R}^{\text{in}}$ and two outgoing modes 
$k_L^{\text{out}}$, $k_{\bar L}^{\text{out}}$. Thus we have three ingoing modes and three outgoing modes and they form a complete basis of the asymptotic modes, respectively. Hence there must exist a linear transformation between them described by a $3\times 3$ matrix, which is the Bogoliubov transformation in the present case and its entries are the Bogoliubov coefficients. 
However, since the expression in the previous section is not valid near the horizon and we have not identified decaying and growing modes, we cannot rely on the matching condition used, for example, in \cite{Fabbri:2010tj,Osawa:2022vkw}. 

In order to bypass this problem, a nice idea is presented in \cite{Coutant:2017nea}, which we follow in this section. 
The point in \cite{Coutant:2017nea} is that we focus on low energy modes with 
$\omega\ll c_s/L$ with $L$ a scale where $v(x)$ and $\rho_0(x)$ changes. 
For example, in the case of \eqref{eq:profile}, $L\simeq \epsilon$. 
Then there exists a region satisfying $L\ll|x|\ll c_s/\omega$. There we can deduce 
the asymptotic form of the solution to the BdG equation perturbatively 
and the existence of the above region justifies its analytic continuation to the asymptotic region where we have the mode functions explicitly. 
Thus we can read how the solutions to the BdG equation can be written in terms of mode functions in this region. 
Then as in the case of the one-dimensional Schr\"{o}dinger equation, we consider a scattering process, which is described by the Bogoliubov transformation between mode functions in the asymptotic region. Since we already know a relation between 
the mode functions and solutions to the BdG equation, we get the Bogoliubov coefficients in terms of parameters of the latter. 

We thus begin with reviewing the results in \cite{Coutant:2017nea} which are relevant for our study. As we mentioned above, it is convenient to 
consider modes with fixed $0<\omega<\omega_c$. In such a case, 
as pointed out in \cite{Macher:2009nz}, it is better to change the normalization of modes accordingly. 

By introducing fields with fixed $\omega$ for $\wh{\delta\phi}$ in \eqref{eq:hat}
\[
\theta_{\omega}&=\frac{1}{2i}\left(\wh{\delta\phi}_{\omega}-\wh{\delta\phi}_{\omega}^{\dagger}\right), \nn \\
n_{\omega}&=\frac{1}{2}\left(\wh{\delta\phi}_{\omega}+\wh{\delta\phi}_{\omega}^{\dagger}\right),
\label{eq:defoftheta}
\]
the BdG equation \eqref{eq:Im}, \eqref{eq:Re} becomes 
\[
\left(-i\omega+v\partial_x\right)\theta_{\omega}
=&\frac{1}{2m\rho_0}\partial_x\left(\rho_0\partial_xn_{\omega}\right)-2\lambda\rho_0n_{\omega}, \nn \\
\left(-i\omega+v\partial_x\right)n_{\omega}
=&-\frac{1}{2m\rho_0}\partial_x\left(\rho_0\partial_x\theta_{\omega}\right).
\label{eq:newBdG}
\]
It is easy to confirm that in the asymptotic region where $v$ and $\rho_0$ are constant, 
we have plane-wave solutions. More precisely, from the result of the previous section, 
they are given as 
\[
x\gg&L: \nn \\
&\theta_R^{\text{out}}=U_re^{ik_R^{\text{out}}(\omega)x}, \nn \\
&\theta_L^{\text{in}}=U_re^{ik_L^{\text{in}}(\omega)x}, \nn \\
 \nn \\
x\ll&-L: \nn \\
&\theta_R^{\text{in}}=U_ee^{ik_R^{\text{in}}(\omega)x}, \nn \\
&\theta_{\bar R}^{\text{in}}=U_ee^{-ik_{\bar R}^{\text{in}}(-\omega)x}, \nn \\
&\theta_L^{\text{out}}=U_le^{ik_L^{\text{out}}(\omega)x}, \nn \\
&\theta_{\bar L}^{\text{out}}=U_le^{-ik_{\bar L}^{\text{out}}(-\omega)x}, 
\label{eq:asymptoticmodes}
\]
where the momenta are shown in \eqref{eq:momenta} and the normalizations are 
\[
U_r=\sqrt{\frac{mc_s}{2\omega\rho_r}}, \qquad U_l=\sqrt{\frac{mc_s}{2\omega\rho_l}}, 
\qquad U_e=\sqrt{\frac{2m^3v_l^2}{\rho_lq_l^3}}
\label{eq:Udef}
\]
with 
\[
q_r=2m\sqrt{c_s^2-v_r^2}, \qquad 
q_l=2m\sqrt{v_l^2-c_s^2},
\label{eq:qdef}
\]
and $\rho_r=-\alpha/v_r$ and $\rho_l=-\alpha/v_l$ are $\rho_0$ in the $x>0$ and $x<0$ region, respectively, as we mentioned below eq.\eqref{eq:theta0}. 

Next let us consider the BdG equation \eqref{eq:newBdG} in $L\ll|x|\ll c_s/\omega$. 
In this region, $\partial_x\sim k\sim 1/|x|\gg \omega/c_s\simeq \omega/v$ 
and hence we can neglect $\omega$ compared to $v\partial_x$. 
Then by using $|x|\gg L$ where $\rho_0$ is constant, 
we obtain an asymptotic form of a solution to the BdG equation given in \cite{Coutant:2017nea} as 
\[
\theta^{(1)}_{\omega}\rightarrow
\begin{dcases}
\frac{1}{iq_l\rho_l}\left(A_2e^{iq_lx}-A_3e^{-iq_lx}\right) & x\ll -L \\
A_1+A_4x+{\cal O}\left(e^{-q_lx}\right) & x\gg L 
\end{dcases}.
\label{eq:theta1}
\]
Here $A_1$,$A_2$,$A_3$ are constants of ${\cal O}(\omega^0)$, while $A_4$ is of ${\cal O}(\omega)$. In order to find them, in \cite{Coutant:2017nea}, by solving the BdG equation up to ${\cal O}(\omega)$ iteratively, we obtain
\[
\frac{v^2-c^2}{v}\partial_x\theta_{\omega}=i\omega\theta_{\omega}
+i\omega c_s^2\int_{-\infty}^{\infty}\frac{\partial_x\theta_{\omega}}{v^2}dx+\frac{1}{2m\rho}\partial_x\left(\rho\partial_xn_{\omega}\right)+{\cal O}(\omega^2),
\]
which holds both on the left and right region. Using this, $A_4$ is found as 
\[
A_4=i\omega\frac{\ve^2+c_s^2}{v_r^2-c_s^2}\frac{v_r}{\ve^2}A_1,
\]
where 
\[
\frac{1}{\ve^2}\equiv\frac{1}{A_1}\int_{-\infty}^{\infty}\frac{\left.\partial_x\theta_{\omega}^{(1)}\right|_{\omega=0}}{v_r^2}dx.
\]
Let us compare eq.\eqref{eq:theta1} and a prediction of the Bogoliubov transformation. As we mentioned at the beginning of this subsection, there must exist a linear transformation between the in-going and out-going modes as\footnote{We follow the parametrization of Bogoliubov coefficients in \cite{Coutant:2017nea}.} 
\[
\begin{pmatrix}
\theta_R^{\text{out}} \\
\theta_{\bar L}^{\text{out}} \\
\theta_L^{\text{out}}
\end{pmatrix}
=S^{-1}
\begin{pmatrix}
\theta_R^{\text{in}} \\
\theta_{\bar R}^{\text{in}} \\
\theta_L^{\text{in}}
\end{pmatrix}
\qquad
\text{with}
\quad
S^{-1}=
\begin{pmatrix}
\alpha^* & -\beta^* & R^* \\
-\tilde{\beta}^* & \tilde{\alpha}^* & -B^* \\
\tilde{R}^* & -\tilde{B}^* & \tilde{T}^*
\end{pmatrix}
.
\label{eq:smatrix}
\]
{}From \eqref{eq:theta1}, we find that $\theta_{\omega}^{(1)}$ is nonzero 
in the asymptotic region on the right ($x>0$), where we have only one outgoing mode $\theta_R^{\text{out}}$. Hence as in the analysis of reflection and transmission of wave in the one-dimensional Schr\"{o}dinger equation, we consider a scattering process on a boundary condition that $\theta_R^{\text{out}}$ is the unique outgoing mode. Then eq.\eqref{eq:smatrix} 
implies that in $x\rightarrow\infty$, we have $\theta_R^{\text{out}}+R^*\theta_L^{\text{in}}$ and that in $x\rightarrow -\infty$, we have $\alpha^*\theta_R^{\text{in}}-\beta^*\theta_{\bar R}^{\text{in}}$. 
Note that this result can be applied to $L\ll |x|$ region where 
$\rho_0(x)$, $v_l$, and $v_r$ are constant. Now we consider the region 
$L\ll |x|\ll c_s/\omega$ and compare this with \eqref{eq:theta1}. 
Here from \eqref{eq:momenta}, for $k=k_R^{\text{out}}, k_L^{\text{in}}, k_L^{\text{out}}, k_{\bar L}^{\text{out}}$, $k|x|\simeq \omega|x|/c_s\ll 1$ and hence for these plane waves in \eqref{eq:asymptoticmodes}, 
$e^{ikx}\simeq 1+ikx$. We therefore have 
\[
\theta\equiv
\begin{dcases}
U_e\left(\alpha^*e^{ik_R^{\text{in}}x}-\beta^*e^{-ik_{\bar R}^{\text{in}}x}\right)
& -\frac{c_s}{\omega}\ll x\ll -L \\
U_r\left(1+R^*+i\left(k_R^{\text{out}}+R^*k_L^{\text{in}}\right)x\right)
& L\ll x\ll \frac{c_s}{\omega} 
\end{dcases}.
\label{eq:theta}
\]
Comparing \eqref{eq:theta1} and \eqref{eq:theta}, we are tempted to identify\footnote
{Here we note that 
$\frac{\omega v_l}{v_l^2-c_s^2}x
\simeq \frac{\omega}{c_s}x\ll 1$ 
and hence $k_R^{\text{in}}x\simeq q_lx$, 
$k_{\bar R}^{\text{in}}x\simeq q_lx$.
} 
them up to normalization constant, i.e. ${\cal N}\theta^{(1)}=\theta$. As shown in \cite{Coutant:2017nea}, this gives some of the Bogoliubov coefficients
\[
R=\frac{\ve^2+c_sv_r}{\ve^2-c_sv_r}, \quad
\alpha=i\frac{A_2U_r}{q_l\rho_lU_eA_1}\left(1+R\right), \quad
\beta=i\frac{A_3U_r}{q_l\rho_lU_eA_1}\left(1+R\right).
\label{eq:Rresult}
\]
At this stage this identification is an ansatz of the connection between the solution to the BdG equation and the asymptotic mode function. Hence consistency should be confirmed by considering another scattering process as done in \cite{Coutant:2017nea}. Thus we consider a scattering process with another boundary condition that the unique outgoing mode is $\theta_{\bar L}^{\text{out}}$. Then \eqref{eq:smatrix} provides asymptotic mode functions $-B^*\theta_L^{\text{in}}$ 
in $x\gg L$, and $\theta_{\bar L}^{\text{out}}-\tilde{\beta}^*\theta_R^{\text{in}}+\tilde{\alpha}^*\theta_{\bar R}^{\text{in}}$ in $x\ll -L$. Accordingly, we consider the solution of the BdG equation as 
\[
\theta^{(2)}=U_l\left(1+i\omega\epsilon(x)\right), 
\]
and we identify ${\cal N}^{(1)'}\theta^{(1)}+\theta^{(2)}$ with the above mode function to get 
\[
\epsilon(x)=
\begin{dcases}
-\frac{1}{v_l+c_s}x & -\frac{c_s}{\omega}\ll x\ll -L \nn \\
\frac{v_r}{v_r^2-c_s^2}\left(1-\frac{c_s}{v_l}\right)x & L\ll x\ll \frac{c_s}{\omega}
\end{dcases},
\]
and 
\[
B=\frac{\ve^2+c_sv_l}{\ve^2-c_sv_r}\sqrt{\frac{v_r}{v_l}}, \quad
\tilde\alpha=-i\frac{A_3U_l}{q_l\rho_lU_eA_1}\left(1+\sqrt{\frac{v_r}{v_l}}B\right), \quad
\tilde\beta=-i\frac{A_2U_l}{q_l\rho_lU_eA_1}\left(1+\sqrt{\frac{v_r}{v_l}}B\right).
\label{eq:Bresult}
\]
The final scattering process with the unique outgoing mode $\theta_L^{\text{out}}$ 
similarly leads to 
\[
\tilde T=\frac{\ve^2-c_sv_l}{\ve^2-c_sv_r}\sqrt{\frac{v_r}{v_l}}, \quad
\tilde R=-i\frac{A_2U_l}{q_l\rho_lU_eA_1}\left(1-\sqrt{\frac{v_r}{v_l}}\tilde T\right), \quad
\tilde B=i\frac{A_3U_l}{q_l\rho_lU_eA_1}\left(1-\sqrt{\frac{v_r}{v_l}}\tilde T\right).
\label{eq:Tresult}
\]

\subsection{Entanglement entropy}
\label{subsec:EE}
Based on the Bogoliubov coefficients derived in the previous subsection, we now compute the entanglement entropy of the analog Hawking radiation. 

As we mentioned in the previous subsection, 
we have mode functions in \eqref{eq:asymptoticmodes} with fixed $\omega$ 
which form a complete basis in the asymptotic region. We therefore introduce the creation and annihilation operators as 
\[
\left.\theta(x,t)\right|_{\omega<\omega_c}
=\int_0^{\omega_c}\frac{d\omega}{2\pi}e^{-i\omega t}
\left(a_R^{\text{in}}\theta_R^{\text{in}}(x)+a_{\bar R}^{\text{in}}\theta_{\bar R}^{\text{in}}(x)+a_L^{\text{in}}\theta_L^{\text{in}}\right)+\text{h.c.},
\] 
for $|x|\gg L$ instead of \eqref{eq:ktoomegalowE}. On the other hand, the outgoing modes also form another complete basis and we can introduce the creation and annihilation operators associated with them. However, these two sets of complete bases are related by the Bogoliubov transformation \eqref{eq:smatrix} and thus operators also have a corresponding relation  
\[
&{\cal A}^{\text{in}}={}^tS^{-1}{\cal A}^{\text{out}}
\label{eq:theBogoliubov}
\]
with
\[
{\cal A}^{\text{in}}=
\begin{pmatrix}
a_R^{\text{in}} \\
a_{\bar R}^{\text{in}\dagger} \\
a_L^{\text{in}}
\end{pmatrix},
\quad
{}^tS^{-1}=
\begin{pmatrix}
\alpha^* & -\tilde\beta^* & \tilde R^* \\
-\beta^* & \tilde{\alpha}^* & -\tilde B^* \\
R^* & -B^* & \tilde{T}^*
\end{pmatrix},
\quad
{\cal A}^{\text{out}}=
\begin{pmatrix}
a_R^{\text{out}} \\
a_{\bar L}^{\text{out}\dagger} \\
a_L^{\text{out}}
\end{pmatrix}.
\label{eq:Bogoliubov_transf}
\]
Then the commutation relations  
\[
\eta_{ij}=\left[{\cal A}_i^{\text{in}},{\cal A}_j^{\text{in}\dagger}\right]
=\left({}^tS^{-1}\eta \left({}^tS^{-1}\right)^{\dagger}\right)_{ij},
\label{eq:constraint}
\]
with $\eta=\text{diag}(1,-1,1)$ impose constraints on the Bogoliubov coefficients. For example, its $(1,2)$-, $(2,2)$- $(3,2)$-, and $(3,3)$- components yield 
\[
0=&-\alpha^*\beta+\tilde\beta^*\tilde\alpha-\tilde R^*\tilde B, 
\label{eq:(1,2)} \\
-1=&\left|\beta\right|^2-\left|\tilde{\alpha}\right|^2+\left|\tilde B\right|^2, 
\label{eq:(2,2)} \\
0=&-R^*\beta+B^*\tilde\alpha-\tilde{T}^*\tilde B,
\label{eq:(3,2)} \\
1=&\left|R\right|^2-\left|B\right|^2+\left|\tilde T\right|^2,
\label{eq:(3,3)}
\]
respectively. 
As mentioned in \cite{Coutant:2017nea}, eq.\eqref{eq:(3,3)} is explicitly confirmed 
by using the results in the previous subsection. On the other hand, the right-hand side in eq.\eqref{eq:(2,2)} is of ${\cal O}(1/\omega)$ and it is explicitly checked 
that the right-hand side indeed vanishes at the leading order in our setup \eqref{eq:vdef}, where $\ve=c_s$.\footnote{In \cite{Mayoral:2010un} subleading contributions are explicitly given with different normalization from ours.} Thus in order to check \eqref{eq:(2,2)}, we have to take account of subleading contributions. 
{}From \eqref{eq:(2,2)} we parametrize them without loss of generality as  
\[
\beta=&e^{i\varphi_1}\cos\theta\sinh r, \nn \\
\tilde{\alpha}=&e^{i\varphi_2}\cosh r, \nn \\
\tilde B=&e^{i\varphi_3}\sin\theta\sinh r.
\label{eq:parameters}
\]

In order to compute the entanglement entropy of the Hawking radiation, let us define the density matrix of the vacuum of the ingoing modes
\[
\rhoin\equiv \vac\,\tvac
\]
with 
\[
a_X^{\text{in}}\vac=0 \quad \text{with} \quad 
X=L, \bar R, R.
\label{eq:vacuum}
\]
Since we regard the $x<0$ region where $|v_l|>c_s$ as an analog BH, 
the Hilbert space of the analog Hawking radiation is that of $R_\text{out}$ modes 
in the $x>0$ region as in Figure \ref{fig:modesummary}. Thus the entanglement entropy of the Hawking radiation is given as the von Neumann entropy of the reduced density matrix obtained by taking the trace of $\rhoin$ over the Hilbert space of the outgoing modes inside the BH: 
\[
S_{\text{EE}}\equiv -\tr_{R_\text{out}}\rho^{(r)}_{R_\text{out}}\ln\rho^{(r)}_{R_\text{out}}, \nn \\
\rho^{(r)}_{R_{\text{out}}}\equiv \tr_{L_{\text{out}}\otimes{\bar L}_{\text{out}}}\rhoin.
\label{eq:rhordef}
\]
For our purpose of computing $S_{\text{EE}}$, it is convenient to introduce a different basis of the Hilbert space of the outgoing modes as in \cite{Nambu:2021lix}, 
which makes structure of the space more transparent:  
\[
A_1^{\text{out}}=&\frac{e^{i\varphi_2}}{\sinh r}\left(\beta^*a_R^{\text{out}}+\tilde B^*a_L^{\text{out}}\right), \nn \\
A_2^{\text{out}}=&a_{\bar L}^{\text{out}}, \nn \\
A_3^{\text{out}}=&\frac{1}{\sinh r}\left(\tilde Ba_R^{\text{out}}-\beta a_L^{\text{out}}\right).
\label{eq:Adef}
\]
Obviously, $\left[A_i^{\text{out}},A_j^{\text{out}}\right]=0$, 
$\left[A_i^{\text{out}},A_j^{\text{out}\dagger}\right]=\delta_{ij}$. 
Combining \eqref{eq:theBogoliubov} and \eqref{eq:Adef}, and utilizing 
eqs.\eqref{eq:(1,2)}, \eqref{eq:(2,2)}, and \eqref{eq:(3,2)}, 
eq.\eqref{eq:vacuum} is translated into 
\[
&\left(A_1^{\text{out}}-\tanh r A_2^{\text{out}\dagger}\right)\vac=0, \nn \\
&\left(-\tanh r A_1^{\text{out}\dagger}+A_2^{\text{out}}\right)\vac=0, \nn \\
&A_3^{\text{out}}\vac=0.
\]
As shown in \cite{Nambu:2021lix}, from this equation, we can express $\vac$ in terms of $A_i^{\text{out}}$ ($i=1,2,3$) as a squeezed state: 
\[
\vac=\frac{1}{\cosh r}\sum_{n=0}^{\infty}\tanh^nr
\ket{n}_{A_1^{\text{out}}}\ket{n}_{A_2^{\text{out}}}\ket{0}_{A_3^{\text{out}}}
\label{eq:psi0}
\]
with 
\[
A_i^{\text{out}}\ket{0}_{A_i^\text{out}}=0, \qquad
\ket{n}_{A_i^{\text{out}}}\equiv\frac{1}{\sqrt{n!}}\left(A_i^{\text{out}\dagger}\right)^n\ket{0}_{A_i^{\text{out}}}.  
\]
It is straightforward to find that 
$\rho^{(r)}_{R_{\text{out}}}$ in eq.\eqref{eq:rhordef} is given by 
\[
&\rho^{(r)}_{R_{\text{out}}}=(1-\xi)\sum_{k=0}^{\infty}\xi^k\ket{k}_{R_\text{out}}
{}_{R_\text{out}}\bra{k} \nn \\
\text{with} \quad 
&\xi=\frac{\cos^2\theta\sinh^2r}{\cosh^2r-\sin^2\theta\sinh^2r}
=\frac{\left|\beta\right|^2}{\left|\tilde\alpha\right|^2-\left|\tilde B\right|^2}
=\frac{\left|\beta\right|^2}{\left|\beta\right|^2+1},
\label{eq:rhor}
\]
where we used \eqref{eq:(2,2)}. 
Details are presented in Appendix \ref{app:rhor}. 
Note that $0<\xi<1$. We therefore conclude
\[
S_{\text{EE}}=-\frac{1}{1-\xi}\Bigl(\left(1-\xi\right)\ln\left(1-\xi\right)+\xi\ln\xi\Bigr). 
\label{eq:SEE}
\]
It is easy to check that $S_{\text{EE}}$ is an increasing function of $\xi$, 
and $\xi$ is also increasing in $r$. Since $\theta_R^{\text{out}}$ is identified as the Hawking mode, the rate of creation of the Hawking pairs on the in-vacuum is 
\[
{}_\text{in}\bra{\psi_0}a_R^{\text{out}\dagger}a_R^{\text{out}}\ket{\psi_0}_{\text{in}}
\propto|\beta|^2, 
\label{eq:rate}
\]
where we used $a_R^{\text{out}}=\alpha a_R^{\text{in}}+\beta a_{\bar R}^{\text{in}\dagger}+Ra_L^{\text{in}}$, which is obtained by inverting eq.\eqref{eq:theBogoliubov} with $S=\eta\left(S^{-1}\right)^{\dagger}\eta$ from eq.\eqref{eq:constraint}. 
As argued in \cite{Osawa:2022vkw}, eq.\eqref{eq:parameters} implies that $r$ corresponds to rate of creation of the Hawking pairs, and hence it is expected 
that $S_{\text{EE}}$ increases monotonically with $r$. This is one of the manifestations of the information loss paradox, and we anticipate that 
in our case backreaction would play an essential role in resolving it. In fact, this is the main motivation of this work. From \eqref{eq:rate}, it is reasonable that $S_{\text{EE}}$ is written in terms of $\beta$. We notice that $|\beta|^2$ not only controls the rate of the Hawking pair production, but also is an average occupation number of the Hawking mode $\vev{\hat{N}}=\text{tr}_{R_\text{out}}\left(\rho^{(r)}_{R_{\text{out}}}\hat{N}\right)=|\beta|^2$, for the number operator $\hat{N}$, and as such contributes to the entanglement entropy. However, it should be noticed that $\beta$ and hence $S_{\text{EE}}$ depend not only on the squeezing parameter $r$ but on $\theta$, which parametrizes the rate at which ${\bar R}_{\text{in}}$ mode is converted to $L_{\text{out}}$ one.

\section{Incorporation of backreaction}
\label{sec:backreaction}
In this section we incorporate backreaction and investigate how it affects the entanglement entropy of the analog Hawking radiation in our setup. 

\subsection{Effect on the density and the velocity}
\label{subsec:changeofrhov}
We first clarify the effect of backreaction on $\rho_0$ and $v$. 
{}From eq.\eqref{eq:backreactedGP}, the backreaction $\delta\phi_0\equiv\wt{\phi}_0-\phi_0$ satisfies 
\[
i\partial_t\delta\phi_0=\left(-\frac{1}{2m}\bm\nabla^2+U\right)\delta\phi_0+\lambda\left(2\left|\phi_0\right|^2\delta\phi_0+\phi_0^2\delta\phi_0^{\dagger}+2\phi_0\vev{\delta\phi^{\dagger}\delta\phi}+\phi_0^{\dagger}\vev{\delta\phi^2}\right).
\]
In the case of the stationary condensate in \eqref{eq:back}, we set 
\[
\phi_0(x,t)=e^{-i\mu t}\bar{\phi}_0(x), \qquad \delta\phi_0(x,t)=e^{-i\mu t}\delta\bar\phi_0(x,t),
\]
and the above equation becomes 
\[
i\partial_t\delta\bar\phi_0=\left(-\frac{1}{2m}\bm\nabla^2+U-\mu\right)\delta\bar\phi_0+\lambda\left(2\left|\bar\phi_0\right|^2\delta\bar\phi_0+\bar\phi_0^2\delta\bar\phi_0^{\dagger}+2\bar\phi_0\vev{\delta\phi^{\dagger}\delta\phi}+e^{2i\mu t}\bar\phi_0^{\dagger}\vev{\delta\phi^2}\right).
\label{eq:backreactioneq}
\]
Its general solution can be written as the sum of a particular solution to 
\[
i\partial_t\delta\bar\phi_0=\lambda\left(2\bar\phi_0\vev{\delta\phi^{\dagger}\delta\phi}+e^{2i\mu t}\bar\phi_0^{\dagger}\vev{\delta\phi^2}\right),
\]
and the general solution to the homogeneous equation  
\[
i\partial_t\delta\bar\phi_0=\left(-\frac{1}{2m}\bm\nabla^2+U-\mu\right)\delta\bar\phi_0+\lambda\left(2\left|\bar\phi_0\right|^2\delta\bar\phi_0+\bar\phi_0^2\delta\bar\phi_0^{\dagger}\right).
\label{eq:general}
\]
Since the latter is nothing but the BdG equation in the stationary case and its solution is interpreted as the analog Hawking radiation, let us restrict ourselves to the former: 
\[
\delta\bar\phi_0=-i\lambda t\left(2\bar\phi_0\vev{\delta\bar\phi^{\dagger}\delta\bar\phi}+\bar\phi_0^{\dagger}\vev{\delta\bar\phi^2}\right).
\label{eq:particular}
\]
Here we used eq.\eqref{eq:bdeltaphi} and absorbed a possible integration constant into the original condensate $\bar\phi_0$.  
Then setting $\wt{\phi}_0=\phi_0+\delta\phi_0=\sqrt{\tilde\rho_0}e^{i\tilde\theta_0}$, 
we have up to ${\cal O}(\lambda)$ correction 
\[
\delta\rho_0\equiv&\tilde\rho_0-\rho_0=\lambda^2\rho_0t^2
\left(2\vev{\delta\bar\phi^{\dagger}\delta\bar\phi}+\tilde C\right)^2>0, \nn \\
\delta\theta_0\equiv&\tilde\theta_0-\theta_0=-\left(\frac{\delta\rho_0}{\rho_0}\right)^{\frac12}.
\label{eq:backreactionform}
\]
Here $\tilde C$ is a coefficient of $\vev{\delta\bar\phi^2}$ obtained as follows: eqs.\eqref{eq:modeexp} and \eqref{eq:ansatz} yield as a tree level correlation function 
\[
\vev{\delta\phi^2}=\tilde Ce^{2im\bm v\cdot\bm x}, \qquad \tilde C=-\int\frac{d^dk}{(2\pi)^d}U_kV_k<0.
\label{eq:defofC}
\]
In the derivation of eq.\eqref{eq:backreactionform}, we also used the fact that $\bar\theta_0=mvx$ for $d=1$, since 
$v(x)$ is piecewise constant. It is easy to see that in $d=1$, $\tilde C$ is logarithmically divergent in the IR and hence we need an IR cutoff or regard the $d=1$ theory as an effective theory as we mentioned at the beginning of section \ref{subsec:setup}. 
On the other hand, $\vev{\delta\phi^{\dagger}\delta\phi}$ is a constant in both $t$ 
and $\bm x$ given by 
\[
\vev{\delta\phi^{\dagger}\delta\phi}=\vev{\delta\bar\phi^{\dagger}\delta\bar\phi}=\int\frac{d^dk}{(2\pi)^d}V_k^2>0, 
\]
which is also UV-divergent. Hence we also need a UV cutoff. For our purpose, however, it is enough that both $\tilde C$ and $\vev{\delta\phi^{\dagger}\delta\phi}$ are real assuming existence of a cutoff and then $\delta\rho_0>0$. Since \eqref{eq:particular} is pure imaginary, positivity of $\delta\rho_0$ seems universal irrespective of the precise form of the two-point functions. Here it is essential that the phase factor in \eqref{eq:defofC} cancels. Since $\delta\rho_0$ is piecewise constant in $x$, we see from eq.\eqref{eq:backreactionform} that $\bm v=\bm\nabla\theta_0/m$ remains unchanged 
by the backreaction. It should be noticed that eq.\eqref{eq:backreactionform} is valid only for $t$ with a small absolute value according to the two-point functions, because we have derived it under assumption that $\delta\rho_0$ is small compared to $\rho_0$.    

It is worth pointing out that $\delta\rho_0>0$ is in accordance with our expectation. 
Namely, eq.\eqref{eq:sv} implies that increase of the density gives rise to that of the sound velocity $c_s$, while the flow velocity $v$ remains intact. Then it is apparent that the sonic horizon where $v=-c_s$ initially located at $x=0$ recedes toward negative $x$.\footnote{It should be noticed that in the case of \eqref{eq:profile}, the velocities of the flow and of the sound happen to be related. However, they originally have different origins, from $\theta_0$ and $\rho_0$ as in eq.\eqref{eq:v} and \eqref{eq:sv}. Therefore $\delta\rho_0>0$ and $\delta v=0$ implies that $\delta c_s>0$, while \eqref{eq:profile} does not change. Then it is easy to see that the horizon satisfying $c_s+\delta c_s+v=0$ moves to the left along the $x$-axis.} 
We note that although eq.\eqref{eq:particular} is quite simple, it is in accordance with our naive picture that the pair creation in the BH background triggers the Hawking radiation. In fact,  
it is also anticipated in \cite{Liberati:2020mdr} that the two-point function plays an essential role in a shrink of the horizon.  

\subsection{Effect on the entanglement entropy}
\label{subsec:changeofEE}
We now study how the entanglement entropy of the analog Hawking radiation 
changes in the presence of the backreaction with $\delta\rho_0>0$.  

\begin{figure}[H]
\centering
\centering \includegraphics[height=2.2in]{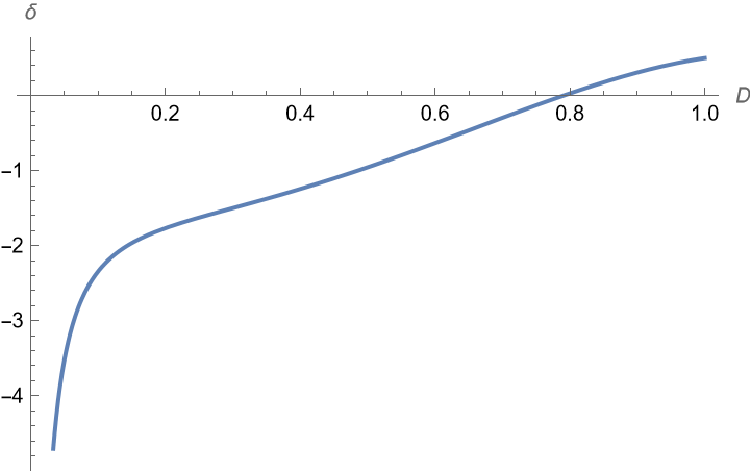}
\vspace{-0cm}
\caption{$\delta$ as a function of $D$ in \eqref{eq:profile}.}
\label{fig:delta}
\end{figure}

{}From eq.\eqref{eq:rhor}  
\[
\frac{1}{\xi}=1+\frac{1}{\left|\beta\right|^2},
\]
and hence decrease of $S_{\text{EE}}$ is equivalent to that of $\left|\beta\right|$. 
Let us therefore examine whether $\delta\rho_0>0$ indeed decreases $\left|\beta\right|$. In the following, physical quantities in the presence of $\delta\rho_0$ are denoted with a tilde. 
Eq.\eqref{eq:backreactionform} implies that 
\[
\delta\rho_0=\lambda^2\rho_0 C'
\]
with $C'>0$ is piecewise constant in $x$, albeit $t$-dependent. 
Here we recall that by construction, $\lambda\rho_0$ and $\rho_0v$ are both constant. Thus we find that $\delta\rho_0$ has different values in the left- ($x<0$) and right- ($x>0$) region because of the $x$-dependence of $\lambda$. Since this difference will be important for later discussion, we note that there exists another constant $C>0$ in $x$ as  
\[
\tilde\rho_{0l}=\rho_{0l}\left(1+v_l^2C\right), \qquad \tilde\rho_{0r}=\rho_{0r}\left(1+v_r^2C\right).
\label{eq:lrdeltarho}
\] 
On the other hand, the sound velocity is defined in \eqref{eq:sv} as a constant since $\lambda\rho_0$ is constant in our setup. 
In the presence of $\delta\rho_{0l}$ and $\delta\rho_{0r}$, however, 
it takes different values in the left- and right-region. We therefore have to distinguish them as 
\[
\tilde c_{sl(r)}=\sqrt{\frac{\lambda\tilde\rho_{0l(r)}}{m}}=c_s
\left(1+\frac12v_{l(r)}^2C+{\cal O}\left(C^2\right)\right).
\label{eq:deltacs}
\]
Accordingly, in eq.\eqref{eq:qdef} $c_s$ in $q_r, q_l$ should be replaced with $c_{sr},c_{sl}$ in $\tilde q_r, \tilde q_l$, respectively. 
Thus we will study how these deformations affect $\beta$. 
For this purpose we need its expression in the case $c_{sl}\neq c_{sr}$. 
Fortunately, in \cite{Coutant:2017nea} the authors provide its explicit form  which is the same as in \eqref{eq:Rresult}, but 
\[
\ve^2=\frac{v_r^2(v_l^2-c_l^2)+v_l^2(c_r^2-v_r^2)}{v_l^2-c_l^2+c_r^2-v_r^2},
\]
which is reduced to $c_s^2$ in our original setup $c_l=c_r=c_s$.  
Then straightforward but tedious calculation shows that 
in the deformation \eqref{eq:lrdeltarho} and \eqref{eq:deltacs}, $|\beta|^2$ 
given by \eqref{eq:Rresult} changes as 
\[
\left|\tilde\beta\right|^2
=&|\beta|^2\left(1+C\delta+{\cal O}(C^2)\right), \nn \\
\delta&\equiv
\frac{1}{4} \left(5v_l^2+5v_r^2-2c_s\left(c_s+2v_r\right)-\frac{3v_l^4}{v_l^2-c_s^2}-\frac{2c_sv_r^3}{c_s^2-v_r^2}-\frac{8v_l^2v_r^2}{c_s^2} \right).
\label{eq:betadeform}
\]
In order to see behaviour of $\delta$ in a typical example, let us concentrate on the well-known case in \eqref{eq:profile}, where $v_l=-c_s(1+D)$, $v_r=-c_s(1-D)$. Then $\delta$ becomes 
\[
\delta=\frac{c_s^2 (2 + 15 D + 34 D^2 - 96 D^3 - 8 D^4 + 55 D^5 - 8 D^7)}{4 D (-4 +D^2)},
\]
and $\delta/c_s^2$ is depicted in Figure \ref{fig:delta}.

As shown in Figure \ref{fig:delta}, $\delta$ is negative over most of the range of $D$ and therefore $|\tilde{\beta}|$ is smaller than $|\beta|$ due to the backreaction as expected, except in a restricted region close to $D=1$. 
There would be several reasons why $\delta$ is not negative for $D\geq 0.8$. The most probable one of them comes from the fact that $v_r$ vanishes for $D=1$. As found in eqs.\eqref{eq:Rresult}, \eqref{eq:Bresult}, \eqref{eq:Tresult}, this is a somehow degenerate case in the Bogoliubov coefficients and then their subleading contribution in the small $\omega$ expansion should be included. Furthermore, our setup in this case requires that $\rho_r$ become divergent and thus $U_r$ in \eqref{eq:Udef} vanishes. Hence the mode functions should be defined more carefully, as with a certain cutoff.

\section{Conclusions and discussions}
\label{sec:discussions}
In this paper we analyzed the effect of backreaction on the entanglement entropy of the analog Hawking radiation in the BEC. We identified the backreaction directly from the modified GP equation, and found that it always increases the density of condensate. This in fact agrees with the expectation that backreaction makes the horizon shrink. We also derived the explicit form of the entanglement entropy in terms of the Bogoliubov coefficients. By using the known results of these coefficients, we explicitly confirmed that the growth of the entanglement entropy is indeed suppressed by the backreaction over most of the parameter region in the well-known example of the analog BH. 

As we mentioned above, in order to clarify what happens near $D=1$, it would be necessary to take account of the next leading order in the small $\omega$ expansion. It seems manageable to do this by, up to this order, considering the dispersion relation as in \cite{Mayoral:2010un}, and solving the BdG equation iteratively. 

In the context of the analog black hole itself, it is important to distinguish universal features from model-dependent ones. For example, we have observed that the backreaction increases the density, and the entanglement entropy is expressed in terms of the Bogoliubov coefficient controlling the rate of creation of the Hawking pairs. These results are expected to be universal and independent of details of models. 

It is meaningful to clarify implications of our results even in the context of the BEC. 
Manifest unitarity convinces us that the Page curve in the BEC must be realized as the time evolution of the entanglement entropy of the Hawking radiation. 
In fact, eq.\eqref{eq:particular} implies that the two-point functions are the source of the backreaction which we have shown reduces the growth of the entanglement entropy. 
Thus our result suggests that the pair creation triggers backreaction, which plays a role of information transfer, as expected. This result should be regarded as a necessary first step toward reproducing the Page curve, rather than a derivation of the Page curve itself. First of all, we have only confirmed that 
the growth of the entanglement entropy is suppressed by a small amount due to the backreaction in quite a short period of time after taking it into account. In order to reproduce the Page curve, we have to keep track of its time evolution at least until the Page time when our approximation of small backreaction cannot be justified. Thus, for example, we have to examine how the effect of backreaction accumulates over time to decrease the total entanglement entropy.  
Second, we computed the entanglement entropy mode by mode with $\omega$ fixed. 
However, the Page curve is the behavior of the entanglement entropy to which all modes contribute. Since we concentrated only on quite low energy modes, our result alone 
does not ensure decrease of such Page curve with all modes.   
Nevertheless from our study at least we can say that the backreaction is one of the most promising candidates responsible for the realization of the Page curve. 
Of course this is what we have expected, but it would be worthwhile to show how to incorporate it and evaluate its effect in an explicit manner as the first step towards our goal.

{}From our original motivation, one of the most important aims is comparison to the genuine BH physics. Of course we understand the relation between the BEC and the BH is not a duality but an analogy. It is, however, important to identify to what extent the analogy persists.  
In the island picture of the gravity side, we have a nonperturbative saddle in the gravitational path integral in calculating the entanglement entropy of the Hawking radiation. 
In particular, the replica wormhole is taken into account as such a saddle in the replica method, and it is known to give the same result as the island formula \cite{Almheiri:2020cfm,Hirano:2021rzg}. 
We note that the existence of the replica wormhole is quite peculiar to quantum gravity, because 
there we need to sum up all allowed space-times on a fixed boundary condition. On the other hand, in our study of the BEC, we do not need a nonperturbative saddle at least for demonstrating the reduction of the growth of the entanglement entropy. 
It is likely that in the BEC side the Page curve will be reproduced without invoking such a nonperturbative solution because it is peculiar to the gravitational path integral as we mentioned above. 
It should be possible to apply the replica method to the BEC as well. Then we prepare copies of the BEC and probably consider the multiple scattering processes according to our computation. It would be quite interesting if we have a nontrivial solution to e.g. GP equation or the BdG equation which appears only in the replica method. 
In the BH case, the Page time is proportional to the Bekenstein-Hawking entropy, which is not available in the BEC case. Furthermore, if there is no IR cutoff, the total entropy in the left region of the horizon is formally divergent. This would suggest that the IR cutoff is necessary for well-definedness of the Page time in the BEC setup not only for the lower-dimension but for the finiteness of the total entropy as in the BH.

In the BEC side, by taking advantage of the existence of the microscopic Hamiltonian, we can compute several important physical observables, for example, two-point functions such as the density-density correlation function. Although it is not straightforward to compute analogous quantities in the gravity side, it would be possible to predict their qualitative behavior by physical arguments in some cases. Furthermore, if the AdS/CFT correspondence is available, or a certain lower-dimensional explicit description like JT gravity can be applied, it is possible to compute their counterparts directly even in the gravity side, by using CFT or a matrix model. In our setup, it is evident that we have separate Hilbert spaces in the left- and the right-region. This would not be the case in the genuine BH from the point of view of non-separability of the Hilbert space inside/outside BH, claimed e.g. in \cite{Geng:2026asi}. According to the central dogma \cite{Almheiri:2020cfm}, however, important physical quantities like the Bogoliubov coefficients can be described in terms of ordinary unitary quantum systems, including standard quantum field theories, at least for an asymptotic observer. In this sense our approach based on the scattering process would be fit for the central dogma. It is expected that our analyses would help distinguish genuinely gravitational features from those common to analog gravity. This distinction would be useful for understanding what quantum gravity is, or resolving the information loss problem.


\section*{Acknowledgments}
We would like to thank Takayuki Hirayama and Shunya Ori for their collaboration at an early stage of this work.
The work of TK was partially supported by JSPS KAKENHI Grant Number 22K03606.


\appendix

\section{Derivation of reduced density matrix} 
\label{app:rhor}

In this appendix details of the derivation of the reduced density matrix \eqref{eq:rhor} are presented. 

{}From the expression in eq.\eqref{eq:psi0}, we first have
\[
\rho^{(r)}_{R_\text{out},L_{\text{out}}}\equiv\text{tr}_{\bar L_{\text{out}}}\rho^{\text{in}}
=\frac{1}{\cosh^2r}\sum_{n=0}^{\infty}\tanh^{2n}r\ket{n}_{A_1^{\text{out}}} {}_{A_1^{\text{out}}}\bra{n}\ket{0}_{A_3^{\text{out}}}{}_{A_3^{\text{out}}}\bra{0}
\]
Since $A_1^{\text{out}}$ and $A_3^{\text{out}}$ are just linear combinations of $a_R^{\text{out}}$ and $a_L^{\text{out}}$, we have 
\[
\ket{0}_{A_1^{\text{out}}}\ket{0}_{A_3^{\text{out}}}=\ket{0}_{R_{\text{out}}}\ket{0}_{L_{\text{out}}}
\]
up to a possible phase factor, which does not appear in the reduced density matrix and hence can be set to 1. 
Then \eqref{eq:Adef} leads to  
\[
\ket{n}_{A_1^{\text{out}}} \ket{0}_{A_3^{\text{out}}}
=\frac{1}{\sqrt{n!}}\left(\frac{e^{i\varphi_2}}{\sinh r}\right)^n
\sum_{k=0}^n\C{n}{k}\beta^k\tilde{B}^{n-k}
\sqrt{k!}\ket{k}_{R_{\text{out}}}\sqrt{(n-k)!}\ket{n-k}_{L_{\text{out}}}.
\]
Applying this result, we obtain 
\[
\rho^{(r)}_{R_{\text{out}}}&=\sum_{m=0}^{\infty}
{}_{L_{\text{out}}}\bra{m}
\rho^{(r)}_{R_\text{out},L_{\text{out}}}
\ket{m}_{L_{\text{out}}} \nn \\
=&\sum_{n=0}^{\infty}\frac{1}{\cosh^{2n+2}r}
\sum_{k=0}^n\C{n}{k}|\beta|^{2k}\left|\tilde{B}\right|^{2n-2k}
\ket{k}_{R_{\text{out}}} {}_{R_{\text{out}}}\bra{k} \nn \\
=&\sum_{k=0}^{\infty}\sum_{l=0}^{\infty}\frac{1}{\cosh^{2k+2l+2}r}
\C{l+k}{k}|\beta|^{2k}\left|\tilde{B}\right|^{2l}
\ket{k}_{R_{\text{out}}} {}_{R_{\text{out}}}\bra{k},
\]
where in the last line we change the order of the sums. 
Taking the sum over $l$ using 
\[
(1-x)^{-k-1}=\sum_{l=0}^{\infty}\C{l+k}{k}x^l,
\]
yields \eqref{eq:rhor}.

\end{document}